\documentclass[10pt,a4paper]{article}
\pdfoutput=1
\usepackage[english]{babel}
\usepackage[latin1]{inputenc}
\usepackage{amsfonts,amsbsy,bm,euscript,mathrsfs}
\usepackage{amssymb,stmaryrd,faktor,slashed}
\usepackage{color}
\usepackage[tbtags]{amsmath}
\usepackage[bookmarks=true,colorlinks=true,linkcolor=black,citecolor=black,urlcolor=black,bookmarksnumbered]{hyperref}
\usepackage[nosort]{cite}

\numberwithin{equation}{section}

\makeatletter
\renewcommand\section{\@startsection {section}{1}{\z@}
{-3.5ex \@plus -1ex \@minus -.2ex}
{2.3ex \@plus.2ex}
{\normalfont\Large\bfseries}}
\renewcommand\subsection{\@startsection{subsection}{2}{\z@}
{-3.25ex\@plus -1ex \@minus -.2ex}
{1.5ex \@plus.2ex}
{\normalfont\large\bfseries}}
\makeatother

\newcommand{\arxivlink}[1]{\href{http://arxiv.org/abs/#1}{arXiv:#1}}
\newcommand \foot [1] {\footnote{#1\vspace{2pt}}}
\newcommand \rf [1] {(\ref{#1})}

\def \texpdf {\texorpdfstring}
\def \be {\begin{eqnarray}}
\def \ee {\end{eqnarray}}

\def \Tr {{\rm Tr}}
\def \STr {{\rm STr}}

\def \bi{\bibitem}
\def \tr {{\rm tr}}
\def \str {{\rm str}}
\def \Tr {{\rm Tr}}

\def \td {\tilde}
\def \ci{\cite}

\def \const {{\rm const}}

\def \a {\alpha}
\def \b {\beta}

\def \del {\partial}
\def \a {\alpha}
\def \aa {{\a'}}
\def \g {\gamma}

\def \ov {\over}

\def \b {\beta}
\def \l {\lambda}

\def \k {\varkappa}

\def \det {\hbox{det}}
\def \ci {\cite}

\def \Tr {{\rm Tr}}

\def \l {\lambda}
\def \const {{\rm const}}

\def \td {\tilde}

\def \bi{\bibitem}
\def \la {\label}
\def \l {\lambda}

\def \ov {\over}

\def \r {\rho}
\def \no {\nonumber}

\def \del {\partial}

\def \bi {\bibitem}
\def \la {\label}
\def \l {\lambda}

\def \r {\rho}

\def \ov {\over}

\def \varpi {{\rm w}}

\def \Tr {{\rm Tr}}

\def \vp {\varphi}

\def \eqref {\rf}
\def \ads {$AdS_3 \times S^3$\ }

\def \iffa {\iffalse}

\def \ads {$AdS_5 \times S^5$\ }

\def \emo {$\eta$-model\ }
\def \lmo {$\l$-model\ }
\def \ff {{\rm f}\,}
\def \gg {{\rm g}\,}
\def \vv {{\rm v}\,}
\def \vk {\varkappa}
\def \hh {h}
\def \Ad {{\rm Ad}}
\def \ddss {{\oplus}}
\def \aa {{\rm a}}
\def \bb {{\rm b}}
\def \hhh {{\rm T}}
\def \xxp {x^-}
\def \xxm {x^+}

\begin{document}

\setcounter{equation}{0}
\setcounter{footnote}{0}
\setcounter{section}{0}

\vspace{-3cm}
\thispagestyle{empty}
\vspace{-1cm}

\rightline{ Imperial-TP-AT-2015-02}
\rightline{ HU-EP-15/21}

\begin{center}
\vspace{1.5cm}

{\Large\bf On integrable deformations of superstring sigma models \\ \vspace{0.2cm} related to $AdS_n \times S^n$ supercosets }

\vspace{1.5cm}

{B. Hoare$^{a,}$\footnote{ben.hoare@physik.hu-berlin.de} and A.A. Tseytlin$^{b,}$\footnote{Also at Lebedev Institute, Moscow. tseytlin@imperial.ac.uk }}\\

\vskip 0.6cm

{\em $^{a}$ Institut f\"{u}r Physik und IRIS Adlershof, Humboldt-Universit\"at zu Berlin,
\\ Zum Gro\ss en Windkanal 6, 12489, Berlin, Germany.}

\vskip 0.3cm

{\em $^{b}$ The Blackett Laboratory, Imperial College, London SW7 2AZ, U.K.}

\vspace{.2cm}
\end{center}

\begin{abstract}
We consider two integrable deformations of 2d sigma models on supercosets
associated with $AdS_n \times S^n$. The first, the ``$\eta$-deformation''
(based on the Yang-Baxter sigma model), is a one-parameter generalization of
the standard superstring action on $AdS_n \times S^n$, while the second, the
``$\l$-deformation'' (based on the deformed gauged WZW model), is a
generalization of the non-abelian T-dual of the $AdS_n \times S^n$ superstring.
We show that the $\eta$-deformed model may be obtained from the $\l$-deformed
one by a special scaling limit and analytic continuation in coordinates
combined with a particular identification of the parameters of the two models.
The relation between the couplings and deformation parameters is consistent
with the interpretation of the first model as a real quantum deformation and
the second as a root of unity quantum deformation. For the $AdS_2 \times S^2$
case we then explore the effect of this limit on the supergravity background
associated to the $\l$-deformed model. We also suggest that the two models may
form a dual Poisson-Lie pair and provide direct evidence for this in the case
of the integrable deformations of the coset associated with $S^2$.
\end{abstract}

\newpage
\setcounter{equation}{0}
\setcounter{footnote}{0}
\setcounter{section}{0}

\tableofcontents

\section{Introduction}\label{secint}

Recently there has been significant interest in two special integrable models
that are closely associated with the superstring sigma model on $AdS_n \times
S^n$. First, in \ci{dmv} a particular integrable deformation of the \ads
supercoset model was considered, generalizing the bosonic Yang-Baxter sigma
model of \ci{Klimcik:2002zj,Klimcik:2008eq,Delduc:2013fga}. Second, in
\ci{hms1,hms2} (generalizing the bosonic model of \ci{Sfetsos:2013wia}) an
integrable model based on the $\hat F/\hat F$ gauged WZW model was constructed,
which is also closely associated to the \ads supercoset. The latter model may
be interpreted as an integrable deformation of the non-abelian T-dual of the
\ads supercoset action.

We shall simply refer to the first model as the ``$\eta$-model'' and to the
second as the ``$\l$-model''. As they contain, as special points, the
original $F/G$ coset model and its non-abelian T-dual model respectively, one
may suspect that they are related by some sort of duality provided one properly
identifies their parameters. Indeed, we shall provide evidence (in the simplest
2d target space case) that they are such a pair of Poisson-Lie dual models
\ci{Klimcik:1995ux,Klimcik:1996np,Alekseev:1995ym,Sfetsos:1999zm,Squellari:2011dg}
hence representing the two ``faces'' of a single interpolating or ``double''
theory.

At the same time, it turns out there is also another, more surprising,
relation: the \emo can be obtained directly from the \lmo as a special limit
(combined with an analytic continuation), which in some sense cuts off the
asymptotically flat region.\foot{This is somewhat similar to how the \ads
background is related to the D3-brane geometry when one decouples the
asymptotic region.} The special point $\k=i$ of the \emo is a pp-wave
background \ci{hrt} that for low-dimensional examples is equivalent in the
light-cone gauge to the Pohlmeyer-reduced (PR) model for the coset theory. This
provides therefore a direct link between the special limit of the \lmo and
the PR model (conjectured in \ci{hms1,hms2} and recently made explicit in
\ci{mira}).

This special limit is of particular interest for understanding the relation of
the \lmo to the $q$-deformation of the light-cone gauge S-matrix
\cite{qsmatrix} for $q$ being a phase. For $q$ real the S-matrix is unitary and
has been shown to be in perturbative agreement \cite{afs,ro} with the \emo of
\cite{dmv}. For $q$ equal a phase, unitarity can be restored \cite{resunit},
and the resulting S-matrix has been conjectured to be related to the \lmo
\cite{hms1,hms2}. However, as the \lmo has no isometries one cannot fix the
associated light-cone gauge and hence there is no apparent connection to the
S-matrix of \cite{qsmatrix}. An important feature of the special limit is that
it generates isometries. It is therefore natural to conjecture that taking an
appropriate limit in the \lmo associated to the $AdS_5 \times S^5$ supercoset
will give the deformed model whose light-cone gauge S-matrix is that of
\cite{resunit}.

\

We shall start in section \ref{s2} with a review of the actions of the
$\eta$-deformed and $\l$-deformed models, considering in detail the relation
between the parameters and also the truncations to the bosonic models.

Then in section \ref{sec2} we shall describe the scaling limit and analytic
continuation that allows one to obtain the metric of the \emo from that of the
$\lambda$-model. We shall discuss the action of this limit on the corresponding
supergravity solution of \ci{Sfetsos:2014cea,Demulder:2015lva} in section
\ref{secsugra} for the models related to the $AdS_2\times S^2$ supercoset.

Finally, in section \ref{sec7} we will conjecture that the two models form a
dual Poisson-Lie pair \ci{Klimcik:1995ux,Klimcik:1996np} and directly verify
this in the case of the integrable deformations of the coset associated to
$S^2$.

In appendix \ref{appb} we shall give different simple forms of the conformally-flat
metrics of the deformed models associated with $S^2$, while in appendix \ref{appdil}
we will discuss an alternative proposal for the dilaton of the models
related to the $AdS_2 \times S^2$ supercoset.

\section{ Deformed models }\label{s2}

\subsection{Supercoset based actions}

We shall consider two integrable 2d models based on the supercosets
\begin{equation}\la{1}
\frac{\widehat{F}}{G_1 \times G_2} \supset \frac{F_1 }{G_1}\times \frac{F_2}{G_2}\ ,
\end{equation}
where $\widehat{F}$ is a supergroup (e.g. $PSU(2,2|4)$ in $AdS_5 \times S^5$ case)
and $F_i$ and $G_i$ are bosonic subgroups.
The superalgebra $\hat{\mathfrak{f}}$ of $\widehat{F}$ admits the
usual $\mathbb{Z}_4$ grading,
with the zero-graded part corresponding to the algebra of $G_1 \times G_2$,
and the bilinear form $\operatorname{STr} = \operatorname{Tr}_{F_1} - \operatorname{Tr}_{F_2}$.

The first ``$\eta$-model'' is defined by the deformed supercoset action of \cite{dmv}
(generalizing the bosonic model of \ci{Klimcik:2002zj})\foot{{We choose Minkowski signature in 2d with $d^2x = dx^0dx^1$ and
$\partial_\pm = \partial_0 \pm \partial_1$.}}
\begin{equation}\label{lagdmv}
\hat I_{h,\eta}(g)= \frac{\hh}{2}\int d^2 x \; \operatorname{STr} \big[g^{-1}\partial_+ g\ P_\eta \frac{1}{1-\frac{2\eta}{1-\eta^2} R_g P_\eta} \ g^{-1}\partial_- g\big] \ ,
\end{equation}
where $g \in \widehat{F}$ and
\begin{equation}
P_\eta = P_2 + \frac{1-\eta^2}{2}(P_1 - P_3) \ , \qquad R_g = \Ad_g^{-1} \, R\, \Ad_g \ . 
\end{equation}
Here $\Ad_g (M) = g M g^{-1}$, $P_r$ are projectors onto the $\mathbb{Z}_4$-graded spaces of $\hat{\mathfrak{f}}$
and the constant matrix $R$ is an antisymmetric solution of the
non-split modified classical Yang-Baxter equation for $\hat{\mathfrak{f}}$.
The overall coupling $\hh$ is the analog of string tension and $\eta$ is the deformation
parameter.\foot{Here the bilinear form $\Tr$ ($\STr$)
is related to the usual matrix trace $\tr$ (supertrace $\str$) by
$\Tr = \nu^{-1}\tr$ for some representation-dependent normalization $\nu$.
We fix this normalization
$\nu$ such that in the undeformed limit $h$ plays the role of the usual string
tension in $AdS_n \times S^n$ backgrounds. In particular, this means that in
the $AdS_2 \times S^2$ case with $\eta = 0$ the bosonic part of the action is
given by
\begin{equation*}
I_{h,0}(g) = \frac{\hh}{2}\int d^2 x \; \big[-(1+\rho^2) \del_+ t \del_- t + \frac{1}{1+\rho^2} \del_+ \rho \del_- \rho
+ (1-r^2)\del_+ \varphi \del_- \vp + \frac{1}{1-r^2} \del_+ r \del_- r \big]\ .
\end{equation*}\label{foot1}}
This action possesses the following $\mathbb{Z}_2$ symmetry:
\begin{equation}\label{z2dmv}
\text{parity} \ , \qquad \hh \to \hh \ , \qquad \eta \to -\eta \ .
\end{equation}
In the undeformed limit, the action \eqref{lagdmv} reduces to the standard supercoset action \cite{Metsaev:1998it,Berkovits:1999zq}
\begin{equation}\label{lagdmvundef}
\hat I_{h,0}(g)= \frac{\hh}{2}\int d^2 x \; \operatorname{STr} \big[g^{-1}\partial_+ g\ P\ g^{-1}\partial_- g\big] \ ,
\qquad \qquad 
P = P_\eta\Big|_{ \eta=0}= P_2 + \frac12 (P_1 - P_3) \ .
\end{equation}
The global $\widehat{F}$ symmetry of this undeformed action is broken by the $\eta$-deformation
to its abelian Cartan subgroup.

The second ``$\l$-model'' \cite{hms2} (generalizing the bosonic model of \ci{Sfetsos:2013wia,tse}) is defined by the action
\be
&&\hat I_{k,\l}(f,A)=\frac{k}{{4}\pi} \Big(
\int d^2 x \; \operatorname{STr}\big[ \frac12 f^{-1}\partial_+ f f^{-1}\partial_- f + A_+ \partial_- f f^{-1}
- A_- f^{-1}\partial_+ f - f^{-1} A_+ f A_- + A_+ A_- \big] \no \\ &&\qquad \qquad \qquad 
- \frac{1}{{3}} \int d^3x \; \epsilon^{abc}\operatorname{STr}
\big[f^{-1}\partial_a f f^{-1}\partial_b f f^{-1}\partial_c f\big]
\label{laghms}
+ (\lambda^{-2} - 1) \int d^2 x \; \operatorname{STr}\big[A_+ P_\lambda A_- \big]\Big)\ ,
\ee
where $f \in \widehat{F}$, \ $A_\pm \in \hat{\mathfrak{f}}$ and
\begin{equation}
P_\lambda = P_2 + \frac1{\lambda^{-1}+1} (P_1 -\lambda P_3) \ .
\end{equation}
The first two lines of \eqref{laghms} correspond to the $ \widehat{F}/\widehat{F}$ gauged WZW model with coupling
(level) $k$ and $\l$ is a deformation parameter.
This action possesses the following
$\mathbb{Z}_2$ symmetry
\begin{equation}\label{z2hms}
\text{parity} \ , \qquad k \to -k \ , \qquad \lambda \to \lambda^{-1} \ ,\qquad A_+ \to \Lambda A_+ \ , \qquad A_- \to \Ad_f(A_- - f^{-1}\partial_- f) \ ,
\end{equation}
where $\Lambda = I + (\lambda^{-2} -1)P_\lambda = P_0 + \lambda^{-2} P_2 + \lambda^{-1} P_1 + \lambda P_3$.

In contrast to \rf{lagdmv} this action has no
global symmetry (there is a $G_1 \times G_2$ gauge
symmetry, which in the end we will always fix).
The interpretation of this action can be understood by considering the special
limit $k\to \infty, \ \l\to 1$ combined with scaling $f\to 1$ as \ci{Sfetsos:2013wia}
\begin{equation}\label{natdlim}
f= \exp(-\frac{{4}\pi}{k}\, v )= 1 - \frac{{4}\pi}{k}v + \mathcal{O}(k^{-2}) \ , \qquad \lambda =
1-\frac{\pi }{k} h + \mathcal{O}(k^{-2}) \ , \qquad k \to \infty \ ,
\end{equation}
where the $\hat{\mathfrak{f}}$ valued field $v$ and the constant $\hh$ are kept fixed in the limit. This
leads to the following action\foot{Note that $\frac{\hh}{2} = \frac{\kappa^2}{4\pi}$, where $\kappa^2$ is the
string tension parameter used in \cite{Sfetsos:2013wia,hms1,hms2} (the definition of $\del_\pm$ used therein had an extra factor of $1/2$ compared to
that used here).}
\begin{equation} \la{9}
\hat I_{k\to \infty,\l\to 1}(f\to 1,A)= \int d^2x \; \operatorname{STr} \big[ \, v \, (\partial_- A_+ - \partial_+ A_- + [A_-,A_+])\big]
+ \frac{\hh}{2} \int d^2 x \; \operatorname{STr} \big[ A_+ P A_- \big]\ ,
\end{equation}
where $P=P_\l \Big|_{\l=1}$ is given in \eqref{lagdmvundef}.
This may be interpreted as a first-order action interpolating between the supercoset action \eqref{lagdmvundef}
(if one first integrates out $v$ giving $A_\pm = g^{-1} \del_\pm g$) and its non-abelian T-dual model (if one first integrates out $A_\pm$).

Thus the meaning of \rf{laghms} is a deformation of the first-order interpolating action \rf{9}.
If one first integrates out $A_\pm$ in \rf{laghms} and gauge-fixes the supergroup element $f$ the resulting sigma model may be viewed as a deformation of the non-abelian T-dual of the original supercoset model \eqref{lagdmvundef}.
At the same time, explicitly integrating out $f$ in \rf{laghms} is not possible in general, so \rf{laghms} does not
apparently have a direct relation to a deformation of the supercoset model \eqref{lagdmvundef}.

While there is a close on-shell connection between the models \eqref{lagdmv}
and \rf{laghms} at the level of classical Hamiltonian (Poisson-bracket)
structures \ci{dmv,hms1,hms2}, establishing their 
correspondence at the level of
the actions (and thus eventually at the quantum level) remains an open problem
that we will attempt to address below.\foot{Note that integrability, together
with expected quantum UV finiteness, suggest that classical relations may in
some way extend to the quantum level.}

\subsection{Relations between parameters}\label{sec1}

Let us now comment on relations between the deformation parameters of the two
models \rf{lagdmv} and \rf{laghms}. The deformation parameters in the two
actions of \cite{dmv} and \cite{hms2} may be defined in terms of the parameter
$\epsilon^2 \in \mathbb{R}$ that appears in the deformed classical Poisson
algebra relations.\foot{For both deformed models, there was a paper focussing
on the bosonic case, \cite{Delduc:2013fga} and \cite{hms1}, written before the
papers discussing the deformation of the superstring, \cite{dmv} and
\cite{hms2} respectively. The parameter $\eta_b$ of \cite{Delduc:2013fga} is
related to the parameter $\eta$ of \cite{dmv} by
\begin{equation*}
\eta_b = \frac{2 \eta}{1-\eta^2} \ ,
\end{equation*}
while the parameter $\lambda_b$ of \cite{hms1} is related to the parameter
$\lambda$ of \cite{hms2} by
\begin{equation*}
\lambda_b = \lambda^2 \ ,
\end{equation*}
To avoid confusion, we will always use the definitions of parameters
as given in the papers discussing the superstring
\cite{dmv,hms2}.}

The relation to the parameter $\eta$ of \cite{dmv} (or $\varkappa$ introduced
in \cite{afs}) is given by
\begin{equation}\label{releps1}
\epsilon^2 = \frac{4 \eta^2}{(1+\eta^2)^2} = \frac{\varkappa^2}{1+\varkappa^2} \ ,\qquad
\qquad \epsilon^2 \in [0,1] \ , \quad \eta^2 \in [0,1] \ , \quad \varkappa^2 \in [0,\infty] \ ,
\end{equation}
where
\begin{equation}\la{211}
\varkappa = \frac{2 \eta}{1-\eta^2} \
\end{equation}
is a natural deformation parameter appearing in the bosonic part of the model \rf{lagdmv}.
Here the ranges describe the deformation considered in
\cite{dmv,afs}.
Note that we could also take
\begin{equation}
\eta^2 \in [1,\infty] \ ,
\end{equation}
to cover the ranges $\epsilon^2 \in [0,1]$ and $\varkappa^2 \in [0,\infty]$. This
is a consequence of the fact that the complex $\eta^2$ plane covers the complex
$\epsilon^2$ and $\varkappa^2$ planes twice. This can be seen explicitly from
the relation
\begin{equation}
\epsilon^2(\eta^2) = \epsilon^2(\frac1{\eta^2}) \ .
\end{equation}
The deformation parameter $\l$ in the action \rf{laghms} of \cite{hms2} is related
to $\epsilon^2$ by
\begin{equation}\begin{split}\label{releps2}
\epsilon^2 = -\frac{(1-\lambda^2)^2}{4\lambda^2} = -\frac{1}{4b^2(1+b^2)}\ ,\qquad
\qquad & \epsilon^2 \in [-\infty,0] \ , \quad \lambda^2 \in [0,1] \ , \quad b^2 \in [0,\infty] \ ,
\end{split}\end{equation}
where we have introduced
\begin{equation}\la{215}
b^2 = \frac{\lambda^2}{1-\lambda^2} \ ,
\end{equation}
which is again a natural deformation parameter in the bosonic part of \rf{laghms}.
Here the ranges describe the deformation considered in \cite{hms2}, but we could also take
\begin{equation}
\lambda^2 \in [1,\infty] \ , \ \ \ \ \qquad b^2 \in [-\infty,-1] \ ,
\end{equation}
to cover the range $\epsilon^2 \in [-\infty,0]$. This is again a consequence of the fact that the complex $\lambda^2$ or $b^2$ planes cover the complex $\epsilon^2$ plane twice, which can be seen explicitly from the relations
\begin{equation}
\epsilon^2(\lambda^2) = \epsilon^2(\frac{1}{\lambda^2})\ ,
\qquad\qquad
\epsilon^2(b^2) = \epsilon^2(-1-b^2)\ .
\end{equation}
For a particular value of $\epsilon^2$ there are four equivalent values of
$\eta$, $b$ and $\lambda$ and two equivalent values of $\varkappa$
as described in the table:
\begin{equation*}
\begin{array}{|c|c||c|c|}
\hline
\eta & -\eta & -\eta^{-1} & \eta^{-1}
\\
\varkappa & - \varkappa & \varkappa & - \varkappa
\\
\lambda & \lambda^{-1} & - \lambda & -\lambda^{-1}
\\
b & \pm\sqrt{-1-b^2} & - b & \mp\sqrt{-1-b^2}
\\ \hline
\end{array}
\end{equation*}
The first and second columns and the third and fourth columns
give rise to equivalent theories in both the two deformations
as they are related by the $\mathbb{Z}_2$ symmetries
\eqref{z2dmv} and \eqref{z2hms}. Furthermore, restricting
to the bosonic models, the first and third columns and
the second and fourth columns give rise to identical deformed theories.
This is a consequence of the fact that
the bosonic truncation of \eqref{lagdmv} depends only on $\varkappa$,
while the bosonic truncation of \eqref{laghms} depends only on $\lambda^2$.

Comparing \rf{releps1} and \rf{releps2} suggests that the parameters of the two deformed models
may be related by an analytic continuation
(choosing signs so that $\l=0,1$ corresponds to $\eta=i,0$) 
\be \la{218}
\eta = i { 1 - \l \ov 1 + \l } \ , \ \ \qquad\qquad \l = {i - \eta \ov i + \eta} \ , \ee
or, equivalently,
\be b^2= - {1 \ov 2} + { i \ov 2\vk} \ , \ \ \ \qquad \ \ \ \ \vk = { i \ov 1 + 2 b^2 } = i\frac{1-\lambda ^2}{1+\lambda ^2} \ . \la{219}\ee
Below we will see that \rf{219} is indeed the relation that allows one to obtain
the \emo \rf{lagdmv} as a special limit (combined with an analytic continuation)
of the \lmo \rf{laghms}.

In addition, this will require us to relate the overall couplings of the two models
by the following analytic continuation (assuming the plus sign in \rf{218})
\be \la{18}
\ \ \ { k \ov \pi} =i {h\ov \vk} \ , \ \ \ \qquad {\rm i.e. } \qquad \ \ \ h= { k \ov \pi ( 1 + 2 b^2) } \ .
\ee
Indeed, \rf{18} is implied by \rf{219} and the expression for $\l$ in \rf{natdlim}, which was required
to obtain the interpolating model \rf{9} for large $k$: with $\l \to 1 - { \pi h \ov k} $ we find from \rf{215}
that $b^2\to {k \ov 2 \pi h} $ and thus, from \rf{219}, that $\vk \to { i \pi h \ov k}$,
in agreement with \rf{18}.

The relation \rf{18} is also consistent with the Pohlmeyer reduction limit, which in the
context of the $\eta$-deformation \cite{dmv} corresponds to taking $\varkappa \to \pm i$,
as discussed in \cite{hrt},
with $h$ being proportional to the level of the underlying $G/H$ gauged WZW model. This then ties in with the Pohlmeyer reduction limit of the deformation of
\cite{hms2} for which $k$ plays the role of the level \cite{mira}.

Remarkably, \rf{18} corresponds to the expected relation between the
quantum deformation parameters $q$ for the two models (cf. \ci{dmv,afs,hms1,hms2}):
\be q= e^{ - {\k \ov h}} \ \ \ \ \ \leftrightarrow \ \ \ \ \ \ q= e^{ - {i\pi \ov k}} \ , \la{19} \ee
with the real $q$ corresponding to the \emo \rf{lagdmv} and the root of unity $q$ to the $\l$-model \rf{laghms}.
Indeed, $q = \exp( -\frac{i\pi}k) $ is the standard expectation for the $q$-deformation
parameter of a WZW type model.

\subsection{Bosonic actions}

It is useful to consider explicitly the bosonic parts of the two models \rf{lagdmv} and \rf{laghms}.
We shall concentrate on the part corresponding to one (compact) $F/G$ factor in \rf{1}.
The bosonic counterpart of the \emo action \rf{lagdmv} is
\be \la{11}
I_{h,\eta} (g) = - { \hh \ov 2} \int d^2 x\ \Tr \big[ J_+ P { 1 \ov 1 - \vk R_g P} J_- \big]\ , \qquad J_a = g^{-1} \del_a g \ , && \\
\vk\equiv {2 \eta \ov 1- \eta^2} \ , \ \qquad \qquad R_g = {\rm Ad}_g^{-1} R\, {\rm Ad}_g \ , && \la{12}
\ee
where $g \in F$, $P=P_2$ is the projector onto the $F/G$ coset part of the algebra ${\mathfrak{f}}$ of $F$
and $R$ is a solution of the modified classical YBE for ${\mathfrak{f}}$.
For $\vk=0$ this becomes the standard $F/G$ coset sigma model.

To make the structure of this action more transparent let us rewrite it in a first-order form. Since
${ 1 \ov 1 - \vk R_g P} = \sum_{n=0}^\infty (\k R_g P)^n$ and $P^2=P$,
introducing an auxiliary field $B_a $ in the coset part of ${\mathfrak{f}}$ (i.e. $P B_a = B_a$) we get
\be -\Tr \big[ J_+ P { 1 \ov 1 - \k R_g P} J_- \big] \ \to \ -\Tr \big[ - B_+ (1 - \k R_g) B_- + B_+ J_- + B_-J_+ \big] \ . \la{13} \ee
Replacing $B_a$ by the field $A_a$ in ${\mathfrak{f}}$, adding a term $A_a C_a$ where $C_a \in \mathfrak{g}$ is in the algebra of $G$ and then redefining
$ {\rm Ad}_g (A_a) = g A_a g^{-1} \to A_a$ we find the following first-order
form of \rf{11}, which has a right-action $G$-gauge symmetry
\be \la{14} I_{h,\eta}
(g, A, C) = -{ h \ov 2} \int d^2 x\ \Tr \big[ - A_+ (1 - \k R) A_- + A_+ D_- g\, g^{-1} + A_- D_+ g\, g^{-1}\big] \ , && \\
D_a g \equiv \del_a g - g C_a \ , \qquad \quad g'= g u\ , \quad C'_a= u^{-1} C_a u + u^{-1} \del_a u \ , \quad u \in G
\ . && \la{15} \ee
This model has parameters $(h,\k)$ and for $ \k\not=0$ the global $F$ symmetry is broken to its Cartan
torus directions.\foot{The canonical choice of $R$ annihilates Cartan generators and
preserves (up to factors) the positive and negative root generators:
$R(T_i)=0, \ R(E_+) = - i E_+, \ R(E_-) = i E_-$.}
In the first-order action \rf{14} the deformation corresponds simply to adding the quadratic $\k A_+ R A_- $ term. Indeed,
we can rewrite \rf{14} as
\be \la{144}
I_{h,\eta} (g, A, C) = -{ h \ov 2} \int d^2 x\ \Tr \big[
D_+ g\, g^{-1}\, D_- g\, g^{-1}
- (A_+ - D_+ g\, g^{-1}) (A_- - D_- g\, g^{-1}) + \k A_+ R A_- \big] \ . \ee
For $\k=0$ one can integrate out $A_a$ giving the standard coset sigma model 
action.\foot{The simplicity of the first-order action \rf{14} is 
related to the simplicity of the corresponding classical Hamiltonian 
description \ci{Delduc:2013fga}. At the same time, its superstring generalization
is not straightforward as $P_\eta$ in \rf{lagdmv} is not a projector and
hence $P_\eta^2 \neq P_\eta$.}

The bosonic part of the \lmo action \rf{laghms} has parameters $(k,\l)$ and a local $G$ symmetry
\be
&&I_{k,\l}(f,A,C)= k \big[ I_{\rm gWZW}(f,A) - { b^{-2} \ov {4}\pi} \int d^2 x\ \Tr (A_a- C_a)^2\big] \ , \ \ \ \ \ \ \ \ \ \ \
b^{-2} \equiv \l^{-2} -1 \ . \la{16}\ \
\ee
Here $f \in F$, $A_a\in {\mathfrak{f}}$ is the gauge field of the $F/F$ gauged
WZW model and $C_a \in \mathfrak{g}$ (the term $(A_a - C_a)^2$ is equivalent to
$(P A_a)^2$). $b$ is a natural deformation parameter (like $\k$ in \rf{12}).
The case of $b \to 0$ corresponds to $A_a=C_a$ or the $F/G$ gauged WZW model.
Another limit is as in \rf{natdlim}, i.e. $k \to \infty$ and $b \to \infty$:
$\l = 1 - { \pi \ov k} h +\ldots$ implies $b^{-2} = { 2\pi \ov k}h +\ldots$.
Then setting $f= 1 - {4\pi \ov k} v + \ldots$ where $v \in {\mathfrak{f}}$ we
find from \rf{16} the bosonic truncation of \rf{9} \ci{Sfetsos:2013wia}
\be
I_{k\to \infty ,\l\to 1} = - \int d^2 x\ \Tr \big[ \, v \, F_{+-}(A) + { h \ov 2} (A_a- C_a)^2\big] \ ,
\la{17}\ \
\ee
where $F_{ab} $ is the field strength of $A_a$. This is the interpolating action for
the $F/G$ coset sigma model and its non-abelian T-dual: if we first integrate over $v$ we get $A_a= g^{-1} \del_a g$, $g \in F$,
and thus the original $F/G$ coset model with tension $h$;
if we first integrate over $A_a$ and $C_a$ we get a sigma model for $v$ which is the non-abelian dual of the $F/G$ coset model.

This suggests that \rf{16} may be viewed as an interpolating model between the
$\lambda$-deformation of the non-abelian T-dual model (a model for the field
$f$ found by first integrating out $A_a$ and $C_a$) and a deformation of the
$F/G$ coset sigma model found by parameterizing $A_a$ in terms of the fields $g$
and $\td g$ (e.g., as $A_a= g^{-1} \del_a g + \epsilon_{ab} \td g^{-1} \del_b
\td g$) and integrating out all fields ($f, \td g, C$) other than $g$. The
latter procedure need not, however, give a local action for $g$ away from the
$k\to \infty, \ b^{-2} \to { 2\pi \ov k}h $ point.\foot{At the same time, since
the deformed \emo action \rf{14} depends not only on the current but also
explicitly on $g$ it does not allow a dualization in an obvious way, i.e. an
analog of a dual model should be non-local.}

While the actions \rf{14} and \rf{16} look very different, having, in particular, different symmetries,
one possibility is that they may be viewed as two dual faces of a ``doubled'' model
related by Poisson-Lie type duality
\ci{Klimcik:1995ux,Klimcik:1996np,Alekseev:1995ym}. The \emo may then be the analog of the ``solvable'' member of the dual pair.
We shall provide
explicit evidence for this in section \ref{sec7} below.

Another possibility to relate the \lmo to the \emo is by a limit that will break the $F/G$ symmetric structure of \rf{16}
to reflect the presence of the $R$-matrix in \rf{11},\rf{14}.
This limit will involve a certain scaling (and analytic continuation) of the group element $f$
plus the map between the parameters \rf{219},\rf{18}.
We shall demonstrate the existence of such limit on various relevant $F/G$ coset examples in the next section.
We shall then study the effect of this limit on the corresponding supergravity backgrounds in section \ref{secsugra}.

\section{ Relating the \texpdf{$\lambda$-model}{lambda-model} to the \texpdf{$\eta$-model}{eta-model} by a limit }\label{sec2}

The target space backgrounds that correspond to the \emo \rf{lagdmv},\rf{11} have abelian isometries
associated to the Cartan directions of the algebra of $F$
that are preserved by $R$-matrix.
At the same time, the
backgrounds that correspond to the \lmo \rf{laghms},\rf{16}
(found by integrating out $A_a$ and fixing a
$G$-gauge on $f$) do not have isometries at all.\foot{This is also a
common feature of backgrounds corresponding to $F/G$ gauged
WZW models with a non-abelian $G$, but for
a non-trivial $\l$ deformation it applies also to the abelian $G$ case \ci{Sfetsos:2013wia}.}
To be able to relate the corresponding metrics we thus need to take a certain scaling limit of the \lmo
in the coordinates corresponding to the Cartan directions of $F$.\foot{The special role of these coordinates may be anticipated from the fact that the \lmo \rf{16} can be viewed as a deformation of the $F/F$ gauged WZW model,
which is a topological theory \ci{spiegel}. In the $F/F$ gauged WZW model
the gauge symmetry ($f'= w^{-1} f w$, \ $w \in F$)
allows one to gauge away all but the Cartan directions, i.e. to choose
$f= e^{ \vp_i T_i}, \ \ [T_i, T_j]=0$, so that the Lagrangian
becomes $
L = \del_+ \vp_i \del_- \vp_i + A_{+ i}\del_- \vp_i - A_{-i} \del_+ \vp_i
$ with $\vp_i = a_i=\const$ as the only solutions.
One may then use these moduli parameters $a_i$ to define certain limits of the deformed
background.}

Below we shall first explicitly demonstrate the existence of such limits on particular
low-dimensional cases, $AdS_2 \times S^2$ and $AdS_3 \times S^3$, and then explain the
general construction for $S^n$
and similar spaces related by analytic continuation.
We shall also explain the relation to the Pohlmeyer reduced model.

\subsection{\texpdf{$AdS_2 \times S^2$}{AdS2 x S2}}\label{sec21}

In the case of $AdS_2 \times S^2$
the relevant bosonic coset space is
\begin{equation}
\frac{SO(1,2)}{SO(1,1)} \times \frac{SO(3)}{SO(2)} \ .
\end{equation}
Starting with the \lmo action \rf{16},
integrating out the gauge field and gauge-fixing the $SO(1,2) \times SO(3)$ field $f$
as\foot{Here $\sigma_i$ are Pauli matrices and $\{(\sigma_1 \ddss 0),(0 \ddss i\sigma_1)\}$ generates the gauge group.
We also take $\Tr = 2 \tr$, where $\tr$ is the usual matrix trace, i.e. $\nu=\frac12$ in footnote \ref{foot1}.}
\begin{equation}\label{patch1}
f= \big[\exp(i t \sigma_3) \exp(\xi \sigma_1)\big] \ddss \big[ \exp(i \varphi \sigma_3) \exp(i\zeta\sigma_1)\big] \ ,
\end{equation}
we find the following metric\foot{We shall use the following notation to relate the bosonic part of the
action to the metric:
$I= \int d^2 x \ G_{mn} (X) \del_+ X^m \del_- X^n $ with
$ds^2 = G_{mn}(X) d X^m d X^n$, i.e. we will absorb all overall constants in the action into the metric.
All the bosonic backgrounds we will consider below will not have a non-trivial
$B$ field
\ci{Grigoriev:2007bu,Demulder:2015lva}. }
\begin{equation}\label{lag1}\begin{split}
2\pi k^{-1} ds^2 = \frac{1}{1+2b^2}\big[ & -dt^2 + \cot^2 t \, d\xi^2
- 4 b^2(1+b^2)(\cosh \xi \, dt - \cot t \sinh \xi \, d\xi)^2
\\
& + d \varphi^2 + \cot^2 \varphi \, d\zeta^2
+ 4 b^2(1+b^2)(\cos \zeta \, d\varphi + \cot \varphi \sin \zeta \, d\zeta)^2 \big] \ .
\end{split}\end{equation}
Note that here, for the $AdS_2$ part, we are considering a different patch of the deformed space
than used in
\cite{Sfetsos:2014cea} which corresponds to
\begin{equation}\label{patch2}
\td f= \big[ \exp(\tilde \xi \sigma_2) \exp(\tilde t \sigma_1)\big]\ddss\big[ \exp(i \varphi \sigma_3) \exp(i\zeta\sigma_1)\big] \ ,
\end{equation}
leading instead to
\begin{equation}\begin{split}\label{lag1a}
2\pi k^{-1}\widetilde {ds}{}^2 = \frac{1}{1+2b^2}\big[ &d\tilde\xi^2 - \coth^2 \tilde \xi \, d\tilde t^2
+ 4 b^2(1+b^2)(\cosh \tilde t \, d\tilde \xi + \coth \tilde \xi \sinh \tilde t \, d\tilde t)^2
\\
& + d \varphi^2 + \cot^2 \varphi \, d\zeta^2
+ 4 b^2(1+b^2)(\cos \zeta \, d\varphi + \cot \varphi \sin \zeta \, d\zeta)^2 \big] \ ,
\end{split}\end{equation}
i.e. related to \eqref{lag1} via the analytic continuation
\begin{equation}
\tilde \xi = i t \ , \qquad \tilde t = \xi \ .
\end{equation}
The reason we consider the patch \rf{lag1} is that it admits a special (singular) field
redefinition
with which we can recover the metric corresponding to the
$\eta$-deformed $AdS_2 \times S^2$ model \cite{Delduc:2013fga,dmv}.

Let us now consider the following (complex) coordinate redefinition
$(t,\xi;\vp,\zeta) \to (t,\rho; \vp, r)$ combined with infinite imaginary shifts of the
$(t,\vp)$ directions (turning them into isometries):
\begin{align}
& t\to t+\frac{i}{2} \log \big[\frac{1-\varkappa^2\rho ^2}{1+ \rho ^2}\big] + i \log \g\ , &&
\xi \to
\frac{1}{2} \log \big[-\frac{1-\varkappa \rho }{1+\varkappa \rho }\big] \ , \nonumber
\\&\varphi \to \varphi+\frac{i}{2} \log \big[\frac{1+\varkappa^2r^2}{1 -r^2}\big] + i \log \g\ , &&
\zeta \to \frac{i}{2}\log\big[-\frac{1+i\varkappa r}{1-i\varkappa r}\big] \ , && \g \to \infty \ .\label{part2}
\end{align}
Here we have introduced the parameter $\k$, which is assumed to be related
to $b$ by \rf{219}. We shall also assume that $k$ is related to $h$ by \rf{18}, i.e.
\be
\la{part1}
b^2 = -\frac{1}{2} + \frac{i}{2 \varkappa } \ , \ \ \qquad \ \ \ \ \ h= \frac{k }{\pi (1+ 2 b^2) }
\ . \ee
Then the metric \rf{lag1} transforms into
\begin{equation}\begin{split}\label{metdmv2}
2 h^{-1} ds^2 = \, & \frac{1}{1-\varkappa^2\rho^2}\big[-(1+\rho^2)dt^2 + \frac{d\rho^2}{1+\rho^2}\big]\\
&\quad+ \frac{1}{1+\varkappa^2 r^2}\big[(1-r^2)d\varphi^2 + \frac{dr^2}{1-r^2}\big] \ ,
\end{split}\end{equation}
i.e. becomes exactly the $\eta$-deformed $AdS_2\times S^2$ metric
\ci{dmv,afs,hrt,Fateev:1992tk}
with $h$ as a tension. Indeed, this metric corresponds to \rf{11} with
$g$ parameterized as
\begin{equation}\begin{split}
g = \big[ \exp (\frac{it}{2} \sigma_3) \exp(\frac12 \operatorname{arcsinh} \rho \, \sigma_2 )\big]\ddss \big[\exp (\frac{i\varphi}{2} \sigma_3) \exp(\frac i2 \operatorname{arcsin} r \, \sigma_2 ) \big]\ ,
\end{split}\end{equation}
and the $R$-matrix chosen to annihilate the Cartan directions $\{i\sigma_3\ddss 0, 0 \ddss i\sigma_3\}$.

This relation between \rf{lag1} and \rf{metdmv2}
involving complex coordinate redefinitions \rf{part2} and a complex map between parameters \rf{part1}
suggests that the \lmo and \emo may correspond to different real ``slices'' of some larger complexified model.

To shed more light on the meaning of the infinite imaginary shift of $t$ and $\vp$ in \eqref{part2}
that plays a central role in the above relation between \rf{lag1} and \rf{metdmv2} it is useful
to repeat the discussion using a simpler (algebraic) choice of coordinates in which the metric becomes conformally flat.
Starting with \rf{lag1} and doing the coordinate redefinition
$(t,\xi; \vp,\zeta) \to (x,y; p,q)$
\begin{align}\nonumber
& t = \operatorname{arccos}\sqrt{x^2-y^2} \ , &
& \xi = \operatorname{arccosh}\frac{x}{\sqrt{x^2-y^2}} \ , & x^2 - y^2 <1\ ,
\\
& \varphi = \operatorname{arccos}\sqrt{p^2+q^2} \ , &
& \zeta = \arccos\frac{p}{\sqrt{p^2+q^2}}\ , & p^2 + q^2 <1 \ , \la{311}
\end{align}
we find
\begin{equation}\la{312}
2 \pi k^{-1} ds^2 = \frac1{1-x^2+y^2}\big[-(1+2b^2) dx^2 + \frac{dy^2}{1+2b^2} \big]
+ \frac1{1-p^2-q^2}\big[(1+2b^2) dp^2 + \frac{dq^2}{1+2b^2}\big] \ .
\end{equation}
Formally continuing to the region for which $x^2 - y^2 > 1$ represents
\eqref{lag1a}, i.e. the original metric of \cite{Sfetsos:2014cea}.
Furthermore, one can check that $x^2 - y^2 = 1$ is a curvature singularity and
hence the two patches covered by \eqref{lag1} and \eqref{lag1a} are separated
by this singularity.

Using again the relation between $(k,b)$ and $(h,\k)$ in \rf{part1} and making an infinite rescaling of the coordinates
\begin{equation}\la{315}
x \to \g \varkappa x \ , \qquad y \to \g y \ , \qquad
p \to \g \varkappa p \ , \qquad q \to \g q \ , \qquad \g \to \infty \ ,
\end{equation}
we get
\begin{equation}\la{316}
2 h^{-1} ds^2 = \frac1{y^2 - \varkappa^2x^2 }\big( dy^2 + dx^2 \big)
+ \frac1{q^2 + \varkappa^2p^2}\big( -dq^2 + dp^2 \big) \ .
\end{equation}
This may be interpreted as the metric of $\eta$-deformed $H^2 \times dS_2$
(euclidean $AdS_2$ times 2d de Sitter space)\foot{The $\k\to \infty $ limit of
\rf{316} gives the same metric as $\k=0$ but with reversed overall sign and the
roles of coordinates interchanged.} background which is related to $AdS_2
\times S^2$ by an analytic continuation.\foot{Note that the ``flat-slicing'' or
Poincar\'e-patch like real coordinates do not exist for $S^2$ but exist for its
analytic continuation $dS_2$.} We will elaborate on this limit (giving its
alternative form) focussing on the $S^2$ part of \rf{lag1} in appendix
\ref{appb}.

The infinite scaling limit \rf{315} relating the \lmo to the \emo amounts to
dropping the constants $1$ in the denominators in \rf{312}. It thus corresponds
to decoupling the asymptotically flat region of the \lmo metric \rf{312} so
that the \emo metric may be interpreted as emerging in a ``near-horizon'' limit
(combined with an analytic continuation of the parameters according to
\rf{219},\rf{18}).

\subsection{\texpdf{$AdS_3 \times S^3$}{AdS3 x S3}}\label{sec3}

Let us now consider the $\l$-deformed action \rf{laghms},\rf{16} for the coset corresponding to
$AdS_3 \times S^3$:
\begin{equation}
\frac{SO(2,2)}{SO(2,1)} \times \frac{SO(4)}{SO(3)} \ .
\end{equation}
Parameterizing the gauge-fixed group-valued field $f$
(for the parts associated to $AdS_3$ and $S^3$ respectively) as
\begin{equation}\begin{split}
f =& \big[\exp(i t (\sigma_3\ddss-\sigma_3)) \exp(\xi (\sigma_1\ddss\sigma_1))\exp(i \psi (\sigma_3\ddss\sigma_3))\big]\\
&\quad\ddss \big[\exp(i \varphi (\sigma_3\ddss-\sigma_3) \exp(i\zeta (\sigma_1\ddss\sigma_1))\exp(i \phi (\sigma_3\ddss\sigma_3))\big] \ ,
\end{split}\end{equation}
and integrating out the
gauge field, we find the following metric (cf. \rf{lag1})\foot{Here we take $\Tr = \tr$, where $\tr$ is the usual matrix trace, i.e. $\nu=1$ in footnote \ref{foot1}.}
\begin{equation}\begin{split}\la{317}
2\pi k^{-1} ds^2 = \frac{1}{1+2b^2}\big[ & - dt^2 + J^2 + \coth^2 \xi\, K^2 - 4b^2(1+b^2)(\cosh^2\xi(dt-K)^2 - J^2)
\\ & +d\varphi^2 + \tilde J^2 + \cot^2 \zeta\, \tilde{K}^2 + 4b^2(1+b^2)(\cos^2\zeta(d\varphi+\tilde{K})^2 + \tilde{J}^2)\big] \ ,
\end{split}\end{equation}
where
\begin{equation}\begin{split}
J = & \csc (2 t) \big( \sin (2 \psi ) d\xi -\coth \xi (\cos (2 t)-\cos (2 \psi ))d\psi \big) \ ,
\\
K = & \csc (2 t) \big( \tanh \xi (\cos (2 t)+\cos (2 \psi ))d\xi - \sin (2 \psi )d\psi\big) \ ,
\\
\tilde{J} = & \csc (2 \varphi ) \big(\sin (2 \phi ) d\zeta +\cot \zeta (\cos (2 \varphi )-\cos (2 \phi )) d\phi\big) \ ,
\\
\tilde{K} = & \csc (2 \varphi ) \big(\tan \zeta (\cos (2 \varphi )+\cos (2 \phi )) d\zeta + \sin (2 \phi ) d\phi \big) \ .
\end{split}\end{equation}
Taking the same limit as in the $AdS_2 \times S^2$ case, i.e. using the redefinitions \eqref{part2} and \eqref{part1},
we find that \rf{317} becomes
\begin{equation}\begin{split}\label{metdmv3}
2 h^{-1}ds^2 = \, &\frac{1}{1-\varkappa^2\rho^2}\big[-(1+\rho^2)dt^2 + \frac{d\rho^2}{1+\rho^2}\big] + \rho^2 d\psi^2\\
&\quad + \frac{1}{1+\varkappa^2 r^2}\big[(1-r^2)d\varphi^2 + \frac{dr^2}{1-r^2}\big] + r^2 d\phi^2 \ .
\end{split}\end{equation}
This is precisely the metric
\ci{afs,hrt,Fateev:1996ea} that corresponds to the deformed $AdS_3 \times S^3$
\emo action \eqref{lagdmv},\rf{11} with $g\in F$ parameterized as
\begin{equation}\begin{split}
g = & \big[\exp (\frac{it}{2} (\sigma_3\ddss-\sigma_3) + \frac{i\psi}2 (\sigma_3\ddss\sigma_3)) \exp(\frac12 \operatorname{arcsinh} \rho \, (\sigma_2\ddss-\sigma_2) ) \big]
\\ & \quad\ddss \big[ \exp (\frac{i\varphi}{2} (\sigma_3\ddss-\sigma_3) + \frac{i\phi}2 (\sigma_3\ddss\sigma_3)) \exp(\frac i2 \operatorname{arcsin} r \, (\sigma_2\ddss-\sigma_2) ) \big]\ ,
\end{split}\end{equation}
and the $R$-matrix chosen to annihilate the
Cartan directions
$\{$$(i\sigma_3\ddss0\ddss 0\ddss0)$, $(0\ddss i\sigma_3 \ddss 0\ddss0)$, $(0\ddss0 \ddss i\sigma_3\ddss0)$, $(0\ddss0 \ddss 0\ddss i\sigma_3) \}$.\foot{Note that there are actually two choices of solution to the corresponding
modified classical YBE, one of which gives the required deformation \eqref{metdmv3} -- see \cite{Hoare:2014oua}.}

\subsection{\texpdf{$S^n$}{Sn} and analytic continuations to \texpdf{$AdS_n$}{AdSn}, \texpdf{$dS_n$}{dSn} and \texpdf{$H^n$}{Hn}}\label{sec41}

Let us now describe a systematic procedure for taking the above limit, relating the
actions of the \lmo and \emo
in the general $AdS_n \times S^n$ case
by considering for simplicity the $F/G$ coset corresponding to the $S^n$ factor, i.e.
\begin{equation}
\frac{SO(n+1)}{SO(n)} \ .
\end{equation}
We shall use the antisymmetric real matrices as the familiar basis of
the algebra $\mathfrak{so}(n+1)$\foot{Here we will also take $\Tr =\frac12 \tr$, where $\tr$ is the usual matrix trace, i.e. $\nu=2$ in footnote \ref{foot1}.}
\begin{equation}\label{basis}
(T_{ab})_{ij} = \delta_{ai}\delta_{bj} - \delta_{aj}\delta_{bi}\ , \qquad a,b,i,j = 1,\ldots, n+1 \ ,
\end{equation}
with the projector onto the coset being given by
\begin{equation}
P_2(M) = - \sum_{a=2}^{n+1} \operatorname{Tr}(M\,T_{1a})\, T_{1a}\ .
\end{equation}
In general, we will choose to parameterize the gauge-fixed
field $f\in F=SO(n+1)$ in the action \rf{16} as
\begin{equation}\label{fff}
f = \exp(2\varphi T_{12})\exp(2\zeta T_{23})\exp(2\phi_1 T_{34}) \exp(2\chi T_{45}) \exp(2\phi_2 T_{56}) \ldots \
\end{equation}
and then take a sequence of limits of the following type
\begin{equation}\label{limit}
\Psi \to \Psi + i \log \g \ , \qquad \qquad \g \to \infty \ ,
\end{equation}
first on $\Psi=\varphi$ and then on every other field in \eqref{fff},
i.e. on $\phi_1$, then on $\phi_2$, etc. This effectively picks
out a Cartan subalgebra of $\mathfrak{so}(n+1)$
\begin{equation}
\{T_{12},\ T_{34},\ T_{56},\ \ldots\}\ ,
\end{equation}
and the angles $\varphi,\phi_1,\phi_2, ...$ will become isometries of the resulting metric.

A couple of comments are in order. First, it is worth noting that for $n$ odd
the last exponential factor in \eqref{fff} is in the sequence and hence
the prescription tells us that we should take the limit in the corresponding field.
In the $S^3$ and $S^5$ examples below this final limit is not necessary:
the previous limits already lead to this direction being an isometry and hence
the limit \eqref{limit} would be trivial (the same should also be true
for all odd $n$).
A related observation is that it always appears to be possible to truncate
easily from $n=2N+1$ to $n = 2N$ by just setting this final angle to zero.
It transpires that to go from $n=2N$ to $n=2N-1$ is not so trivial. This is
not so much to do with taking the limit, rather with
the field redefinitions and analytic continuations that we need to perform
to recover the metrics of \cite{afs,hrt,Lunin:2014tsa}.

In the following
we will consider the two non-trivial cases $n=3$ (already discussed in section \ref{sec3} above) and $n=5$, with the $n=2$
and $n=4$ examples following as simple truncations.
It will be useful to
define the following functions
\begin{equation}\la{vv}
\ff (r) = \frac{1}{1+\varkappa^2 r^2} \ , \ \ \ \ \ \ \qquad \gg (r) = \frac{1}{1-r^2} \ , \ \ \ \ \ \ \qquad \vv(r,\theta)
= \frac{1}{1+\varkappa^2r^4\sin^2 \theta} \ .
\end{equation}

\

$\mathbf{n=3}$ and $\mathbf{n=2}$:
Starting with \rf{16} and taking the limits
as described above we end up with a metric with two isometric
directions $\varphi$ and $\phi_1$. There are then two analytic
continuations/coordinate redefinitions that are of particular interest.
The first is given by
\begin{align}\label{def31}
&
\varphi \to \varphi +\frac{i}{2} \log \big[\frac{1+\varkappa ^2 r^2}{1-r^2}\big]\ , &&
\zeta \to \frac{i}{2} \log \big[-\frac{1+i \varkappa r}{1-i \varkappa r}\big] \ , &&
\phi_1 \to \phi_1 \ ,
\end{align}
and the resulting metric is as in \rf{metdmv3} (with $\phi=\phi_1$)
\begin{equation}\label{met31}
2h^{-1}ds^2 = \ff(\gg^{-1}d\varphi^2 + \gg dr^2) + r^2 d\phi_1\ .
\end{equation}
This metric is precisely the deformation of $S^3$ arising from the corresponding \emo \cite{afs,hrt,Hoare:2014oua,Fateev:1996ea}:
it follows from the
\emo action \eqref{lagdmv},\eqref{11} with $g \in F$ parameterized as
\begin{equation}\label{gso3}
g = \exp(\phi_1 T_{34})\exp(\varphi T_{12}) \exp(\arcsin r T_{13}) \ ,
\end{equation}
and the $R$-matrix chosen to annihilate the Cartan directions $\{T_{12},T_{34}\}$.
The second change of variables is given by
\begin{align}\label{def32}
& \varphi \to i \varkappa \varphi +\frac{i}{2} \log \big[\frac{1-r^2}{1+\varkappa^2 r^2}\big]\ ,
&&
\zeta \to \frac{i}{2} \log \big[\frac{1- r}{1+r}\big] \ ,
&&
\phi_1 \to i\varkappa \phi_1 \ ,
\end{align}
with the resulting metric being
\begin{equation}\label{met32}
2h^{-1}ds^2 =
\gg(\ff^{-1} d\varphi ^2 + \ff dr^2)
+r^{-2}d\phi_1^2\ .
\end{equation}
This metric is related to \rf{met31} by two T-dualities -- in each of the isometric directions $\varphi$ and $\phi_1$.
Furthermore, there is a formal map between the two metrics \eqref{met31} and \eqref{met32} given by
\begin{equation}\label{fm1}
\varphi \to i \varkappa \varphi \ , \qquad r \to \frac{i}{\varkappa r} \ , \qquad \phi_1 \to i \varkappa \phi_1 \ .
\end{equation}

To recover the corresponding expressions for $n=2$ one can consistently truncate by setting $\phi_1 = 0$.

$\mathbf{n=5}$ and $\mathbf{n=4}$:
Taking the limits as
described above, from \rf{16} we find a metric with three isometric directions $\varphi$,
$\phi_1$ and $\phi_2$. There are again two analytic continuations/coordinate redefinitions that are of particular interest.
The first is given by
\begin{align}
&
\varphi \to \varphi +\frac{i}{2} \log \big[\frac{1+\varkappa ^2 r^2}{1-r^2}\big]\ ,
&&
\phi_1 \to i\varkappa \phi_1 + i \log \cos \theta \ ,
&&
\phi_2 \to \phi_2 \ ,
\nonumber \\ &
\label{def51}
\zeta \to \frac{i}{2} \log \big[-\frac{1+i \varkappa r}{1-i \varkappa r}\big] \ ,
&&
\chi \to \frac{i}2 \log\big[-\frac{1-\sin\theta}{1+\sin\theta}\big]\ ,
&&
\end{align}
and the resulting metric is
(with $\ff,\gg,\vv$ defined in \rf{vv})
\begin{equation}\label{met51}
2h^{-1}ds^2 =
\ff(\gg^{-1} d\varphi ^2 + \gg dr^2)
+\frac{(d\phi_1 + \varkappa r^4 \vv \sin \theta\cos\theta\, d\theta)^2}{r^2 \vv \cos^2\theta }
+r^2 \vv d\theta^2
+r^2 \sin ^2\theta\, d\phi_2^2 \ .
\end{equation}
As shown in \cite{Lunin:2014tsa}, this metric is T-dual to the metric
constructed in \cite{afs}, which follows from the $\eta$-model \eqref{lagdmv},\eqref{11}
of \cite{Delduc:2013fga,dmv} with
$g \in F$ parameterized as
\begin{equation}\label{gso5}
g = \exp(\phi_2 T_{56})\exp(\phi_1 T_{34})\exp(\theta T_{35})\exp(\varphi T_{12}) \exp(\arcsin r T_{13}) \ ,
\end{equation}
and the $R$-matrix chosen to annihilate the Cartan directions $\{T_{12},T_{34},T_{56}\}$.
Here the T-duality
should be done in just the $\phi_1$ isometry,
making the metric diagonal but generating
a non-zero $B$-field, in agreement with the background found in \cite{afs}.\foot{Note that all
the $\l$-model
backgrounds corresponding to the choice of $f$ in \rf{fff} have no $B$ field \ci{Grigoriev:2007bu,Demulder:2015lva}.}

The second change of variables is given by
\begin{align}\no
& \varphi \to i \varkappa \varphi +\frac{i}{2} \log \big[\frac{1-r^2}{1+\varkappa^2 r^2}\big]\ ,
&&
\phi_1 \to i\varkappa \phi_1 + i \log \cos \theta \ , &&
\phi_2 \to i \varkappa \phi_2 \ ,
\\\label{def52} &
\zeta \to \frac{i}{2} \log \big[\frac{1- r}{1+r}\big] \ ,
&&
\chi \to \frac{i}2 \log\big[\frac{1-\sin\theta}{1+\sin\theta}\big]\ ,
&&
\end{align}
leading to
\begin{equation}\label{met52}
2h^{-1}ds^2 =
\gg(\ff^{-1} d\varphi ^2 + \ff dr^2)
+\frac{(d\phi_1 + \varkappa r^4 \vv \sin \theta\cos\theta d\theta)^2}{r^2 \vv \cos^2\theta }
+r^2 \vv d\theta^2
+r^{-2} \operatorname{csc}^2\theta\, d\phi_2^2 \ .
\end{equation}
This metric (related to \rf{met51} by two T-dualities) is also T-dual to the metric found in \cite{afs}: here one needs three
T-dualities -- in each of the isometric directions $\varphi$, $\phi_1$
and $\phi_2$.
There is again a formal map between the two metrics \eqref{met51} and \eqref{met52} given by
\begin{equation}\label{fm2}
\varphi \to i \varkappa \varphi \ , \qquad r \to \frac{i}{\varkappa r} \ , \qquad \phi_1 \to \phi_1 + i \log \sin \theta \ ,
\qquad \theta \to i\log\big[-i \tan\frac{\theta}2\big] \ , \qquad \phi_2 \to i \varkappa \phi_2 \ .
\end{equation}

To obtain similar expressions for the $n=4$ case one can consistently
truncate by setting $\phi_2 = 0$ in the $n=5$ expressions.

\

Let us now briefly outline the analytic continuations to $AdS_n$, $dS_n$ and $H^n$.
These geometries are all based on different real forms of the
complexified coset space
$\frac{SO(n+1,\mathbb{C})}{SO(n,\mathbb{C})}$, i.e.
\begin{align}
& S^n = \frac{SO(n+1)}{SO(n)} \ , && AdS_n = \frac{SO(2,n-1)}{SO(1,n-1)}\ ,
&& dS_n = \frac{SO(1,n)}{SO(1,n-1)} \ , && H^n = \frac{SO(1,n)}{SO(n)} \ .
\end{align}
After a brief study of the group elements of interest
\eqref{fff},\eqref{gso3},\eqref{gso5} one can see that for $H^n$ there is
essentially one analytic continuation of the basis \eqref{basis}, while for
$AdS_n$ and $dS_n$ there are many potentially inequivalent ones, which in turn
may lead to metrics covering different coordinate
patches of the \emo and $\lambda$-model metrics.

For $AdS_n$ one choice of analytic continuation is given by
\begin{equation}
T_{1\hat a} \to iT_{1\hat a} \ , \qquad T_{2\hat a} \to iT_{2\hat a}
\ , \qquad \hat{a} = 3,\ldots, n+1 \ .
\end{equation}
for which the subalgebra commuting with $T_{12}$, spanned by $T_{\hat a \hat b}$,
remains $\mathfrak{so}(n-1)$. This corresponds to analytically continuing the fields as follows
\begin{equation}\begin{split}\label{adsac}
& \varphi \to t \ , \qquad \phi_i \to \psi_i \ , \qquad
\zeta \to i \xi \ , \qquad \chi \to \hat \chi \ , \qquad
r \to i \rho \ , \qquad \theta \to \hat \theta \ .
\end{split}\end{equation}
Here we also need to flip the overall sign of the metrics.
Other possible analytic continuations involve $T_{12} \to i T_{12}$, so that
the subalgebra commuting with this generator is then $\mathfrak{so}(1,n-2)$.
It is an analytic continuation of this form that is required to obtain the
first line of \eqref{lag1a} from the second line and was considered in the
supergravity constructions of \cite{Sfetsos:2014cea,Demulder:2015lva}.

For $dS_n$ one choice of the analytic continuation is given by
\begin{equation}
T_{12} \to i T_{12} \ , \qquad T_{2\hat{a}} \to i T_{2\hat{a}} \ , \qquad \hat{a} = 3,\ldots n+1 \ ,
\end{equation}
for which the subalgebra commuting with $T_{12}$, spanned by $T_{\hat a \hat b}$,
remains $\mathfrak{so}(n-1)$. This corresponds to analytically continuing the fields as follows
\begin{equation}\begin{split}\label{dsac}
& \varphi \to i t \ , \qquad \phi_i \to \psi_i \ , \qquad
\zeta \to i \xi \ , \qquad \chi \to \hat \chi \ , \qquad
r \to \rho \ , \qquad \theta \to \hat \theta \ .
\end{split}\end{equation}
The remaining analytic continuations,
which we will not explore in detail here, involve leaving $T_{12}$
as is, so that
the subalgebra commuting with this generator is again $\mathfrak{so}(1,n-2)$.

To recover the coset and deformed models associated to $H^n$
we analytically continue
\begin{equation}\label{hac1}
T_{1\bar a} \to i T_{1\bar a} \ , \qquad \varphi \to i \varphi \ , \qquad r \to i r \ , \qquad
\bar a = 2,\ldots, n+1\ ,
\end{equation}
and, as for $AdS_n$, flip the overall sign of the metrics.
It will also be useful to give the direct analytic continuation of
the fields from $AdS_n$ to $H^n$, i.e. combining the inverse of
\eqref{adsac} and \eqref{hac1}
\begin{equation}\label{hac}
t \to i\varphi \ , \qquad \phi_i \to \psi_i \ , \qquad
\xi \to - i \zeta \ , \qquad \chi \to \hat \chi \ , \qquad
\rho \to r \ , \qquad \theta \to \hat \theta \ .
\end{equation}

\subsection{Relation to the Pohlmeyer-reduced model for \texpdf{$AdS_n \times S^n$}{AdSn x Sn} and 
the \texpdf{$\eta \to i$}{eta->i} / \texpdf{$\l \to 0$}{lambda->0} limit}\label{secpohlmeyer}

The Pohlmeyer-reduced model is conjectured to be related to the
\lmo at the special point in the 
parameter space $\l=0$ or $b=0$ \cite{hms1,hms2}, or, equivalently, according to \rf{219}, 
$\eta=i$ or $\varkappa = i$. For
this point the relation between the overall couplings \eqref{18} becomes $h
=\frac{k}{\pi}$. As discussed beneath \eqref{16} the $b \to 0$ limit of the
\lmo gives the $F/G$ gauged WZW model. On the other hand, it was shown in
\cite{hrt} that for the $\eta$-models arising as deformations of $AdS_2 \times
S^2$ and $AdS_3 \times S^3$ models the $\varkappa \to i$ limit of \eqref{met31} can be
taken in such a way (combining it with a coordinate redefinition) 
 that it gives a string action in a pp-wave type background, whose light-cone gauge-fixing is the
Pohlmeyer reduction (PR)
 \ci{Grigoriev:2007bu,Mikhailov:2007xr,Grigoriev:2008jq} of these $AdS_n \times S^n$ models.\foot{If
one takes the $\k\to i$ limit of the \emo without rescaling the coordinates 
 the resulting action gives the same model without the potential term,
i.e. one time and one space dimension decouple.
The metric in the ``transverse'' directions is that of the 
$SO(2) \times SO(1,1)$ and $\frac{SO(3)}{SO(2)} \times \frac{SO(1,2)}{SO(2)}$ gauged
WZW models for $n=2$ and $n=3$ respectively.}

In section \ref{sec41} we considered a sequence of special coordinate
redefinitions that led from the \lmo to (T-duals) of the $\eta$-model. In the
cases of $S^2$ and $S^3$ there was only one limit in this sequence
\eqref{def31}. 
One can thus
see the emergence
of the PR model from the \lmo
in a special limit (cf. also \ci{hms1,mira}).

In the $AdS_5 \times S^5$ case the $\varkappa \to i$ limit of the \emo did not lead directly to
the PR model,
but rather to a
closely related theory with an imaginary $B$ field \ci{hrt}.
It is now clear that there
is a natural ``intermediate''
candidate model
for recovering the PR
model found by making only the first coordinate redefinition in
the sequence \eqref{limit},\eqref{def51} along with the corresponding one for $AdS_5$
\begin{align}
& t\to t+\frac{i}{2} \log \big[\frac{1-\varkappa^2\rho ^2}{1+ \rho ^2}\big] + i \log \g\ , &&
\xi \to
\frac{1}{2} \log \big[-\frac{1-\varkappa \rho }{1+\varkappa \rho }\big] \ , \nonumber
\\&\varphi \to \varphi+\frac{i}{2} \log \big[\frac{1+\varkappa^2r^2}{1 -r^2}\big] + i \log \g\ , &&
\zeta \to \frac{i}{2}\log\big[-\frac{1+i\varkappa r}{1-i\varkappa r}\big] \ , && \g \to \infty \ , \label{bb}
\end{align}
and using the relation of the parameters in \eqref{part1}.
It is interesting to note that considering the analytic continuation to $H^5 \times dS_5$
given in \eqref{dsac},\eqref{hac} this becomes
\begin{align}\nonumber
&\varphi \to \varphi+\frac{1}{2} \log \big[\frac{1-\varkappa^2r^2}{1 +r^2}\big] + \log \g\ , &&
\zeta \to \frac{i}{2}\log\big[-\frac{1-\varkappa r}{1+\varkappa r}\big] \ ,
\\ \la{bbb}
& t\to t+\frac{1}{2} \log \big[\frac{1+\varkappa^2\rho ^2}{1- \rho ^2}\big] + \log \g\ , &&
\xi \to \frac{1}{2} \log \big[-\frac{1+i\varkappa \rho }{1-i\varkappa \rho }\big] \ , && \g \to \infty \ ,
\end{align}
which for $\varkappa^2 \in (0,-1]$ is a real field redefinition and real limit.
Furthermore, for $\varkappa$ in this range the map between the parameters
\eqref{part1} also becomes real.
Therefore, this limit of the $AdS_5 \times S^5$ \lmo can be thought of as
first an analytic continuation to $H^5 \times dS_5$, then a real limit and
field redefinition and finally analytically continuing back.

Following this procedure we find a somewhat involved metric, which has
isometric directions $t$ and $\varphi$ and importantly is real for $\varkappa^2
\in (0,-1]$.\foot{Recall that if we take the second special limit for $\phi_1$ in \eqref{def51}
the off-diagonal terms in the resulting metric \eqref{met51} are imaginary for this
range of $\varkappa$.} Therefore, it is natural to conjecture that the light-cone
gauge-fixing of this model is related to the kink S-matrix of \cite{resunit}.\foot{This
discussion is also true if we only consider the first coordinate
redefinition in the sequence \eqref{limit},\eqref{def52}, however, the resulting
metrics are diffeomorphic as they are related by the map
\begin{equation*}
t \to i \varkappa t \ , \qquad \rho \to -\frac{i}{\varkappa\rho} \ , \qquad
\varphi \to i \varkappa \varphi \ , \qquad r \to \frac{i}{\varkappa r} \ .
\end{equation*}
which is real for $\varkappa^2 \in (0,-1]$.}

The limit of \cite{hrt}
\begin{equation}\la{111}
t = \epsilon \xxp + \frac{\xxm}{\epsilon} \ , \qquad
\varphi = \epsilon \xxp - \frac{\xxm}{\epsilon} \ , \qquad
\rho = \tan \alpha \ , \quad r = \tanh\beta \ , \quad \varkappa = \sqrt{-1 + \epsilon^2}\ , \quad \epsilon\to 0\ ,
\end{equation}
for the $AdS_3 \times S^3$ \emo gives a pp-wave type model whose light-cone gauge fixing
is the Pohlmeyer reduction of strings on $AdS_3 \times S^3$ \cite{Grigoriev:2008jq}
with axial gauging of the associated gauged WZW model. In higher dimensions
the gauge group of the PR theory is no longer abelian and hence axial gauging is not
possible. Therefore, the limit \eqref{111} needs a mild modification to
extract the vector gauged model
\begin{equation}\la{111a}
t = \epsilon \xxp + \frac{\xxm}{\epsilon} \ , \qquad
\varphi = \epsilon \xxp - \frac{\xxm}{\epsilon} \ , \qquad
\rho = \cot \alpha \ , \quad r = \coth\beta \ , \quad \varkappa = \sqrt{-1 - \epsilon^2}\ , \quad \epsilon\to 0\ .
\end{equation}
Taking this limit
in the model obtained by the special limit \eqref{bb} of the \lmo
associated to $AdS_5 \times S^5$ we find a pp-wave type metric
(recall that in this limit we
get from \rf{18} that $h = \frac{k}{\pi}$)
\begin{equation}\begin{split}
& 2h^{-1} ds^2 = -4d\xxp d\xxm + \frac12 (\cos \alpha -\cosh\beta)\, (d\xxm)^2 + ds_{A\perp}^2(\alpha,\psi_1,\hat \chi,\psi_2) + ds_{S\perp}^2(\beta,\phi_1,\chi,\phi_2)\ ,
\end{split}\end{equation}
where the ``transverse'' metrics $ds_{A\perp}^2$ and $ds_{S\perp}^2$ are those of the gauged WZW model
for $\frac{SO(5)}{SO(4)}$ and $\frac{SO(1,4)}{SO(4)}$ respectively.\foot{We parameterize the gauge-fixed field $f_{_{\rm PR}} \in
SO(5) \times SO(1,4)$ of the PR model as
\begin{equation*}
f_{_{\rm PR}} = \big[\exp(2 \alpha T_{23})\exp(\psi_1 T_{34}) \exp(\hat \chi T_{45}) \exp(\psi_2 T_{56})\big]
\ddss \big[\exp(2 i \beta T_{23})\exp(\phi_1 T_{34}) \exp(\chi T_{45}) \exp(\phi_2 T_{56})\big] \ ,
\end{equation*}
and integrate out the gauge field.}
The light-cone gauge-fixing of this model ($\xxm=\mu \tau$) corresponds therefore to the Pohlmeyer-reduced
theory for strings on $AdS_5 \times S^5$ \cite{Grigoriev:2007bu}. Note that as for the
$AdS_2 \times S^2$ and $AdS_3 \times S^3$ cases, the roles of the $AdS_n$ and $S^n$
are effectively interchanged, i.e. the $\varkappa \to i$ limit of the deformed $AdS_5$ metric
leads to the PR model for the string on $\mathbb{R} \times S^5$ and vice versa.

\section{Supergravity backgrounds for deformed models: \texpdf{$AdS_2 \times S^2$}{AdS2 x S2}} \la{secsugra}

Having discussed the form of the metrics corresponding to the \emo and \lmo let
us now consider their extension to the full type IIB supergravity backgrounds
expected to be associated with the superstring actions \rf{lagdmv} and
\rf{laghms}. The direct construction of such backgrounds supporting the
metrics of \emo turns out to be quite non-trivial \ci{afs,Lunin:2014tsa}. At
the same time, the RR backgrounds supporting the \lmo metrics appear to be
much simpler and they were found explicitly in the $AdS_n \times S^n$ cases in
\ci{Sfetsos:2014cea} ($n=2,3$) and \ci{Demulder:2015lva} ($n=5$).

Given that the metrics of \emo can be obtained, as explained above, from the
metrics of the \lmo by a special scaling limit and analytic continuation, one
may reconstruct the full supergravity backgrounds that emerge when this limit
is applied to the solutions of \ci{Sfetsos:2014cea,Demulder:2015lva}. This will
be explored below on the simplest $AdS_2 \times S^2$ example. Surprisingly, the
resulting limiting background will be different from the one constructed in
\ci{Lunin:2014tsa}, even though the two share the same metric \rf{metdmv2}.
Understanding the proper meaning of this solution (that takes a
very simple form in the algebraic coordinates introduced in \rf{311},\rf{312})
will be left for the future.


To discuss the deformed backgrounds associated with the $AdS_2 \times S^2$ supercoset
it is useful to follow \ci{Lunin:2014tsa} and consider the compactification of 10d type IIB supergravity to four
dimensions on $T^6$ retaining only the metric, dilaton and a single RR 1-form potential
$A=A_m dx^m$.\foot{The corresponding 10d 5-form strength will be expressed in terms of the product of the
2-form $F$
and holomorphic 3-form on $T^6$ as in (A.19) of \ci{Lunin:2014tsa}.}
The resulting bosonic 4d action is then given by
\begin{equation}\label{sact}
\mathcal{S} = \int d^4x \; \sqrt{-g}\Big[e^{-2\Phi}\big[R + 4(\nabla \Phi)^2\big] - \frac14 F_{mn}F^{mn}\Big]\ .
\end{equation}
The corresponding equations of motion are
\begin{equation}\label{seom}
R + 4 \nabla^2 \Phi - 4 (\nabla\Phi)^2 = 0\ , \qquad R_{mn} + 2 \nabla_m\nabla_n \Phi
= \frac{e^{2\Phi}}2 (F_{mp}F_n{}^p - \frac14 g_{mn}F^2) \ , \qquad \partial_n (\sqrt{-g}F^{mn}) = 0 \ .
\end{equation}
The first two equations imply that the dilaton should satisfy $\nabla^2 e^{-2 \Phi} =0$. 

\subsection{Angular coordinates}

Our starting point will be the supergravity solution of \cite{Sfetsos:2014cea}
supporting the \lmo metric \rf{lag1a}\foot{In appendix \ref{appdil} we discuss an
alternative 
 choice of the dilaton 
 based on the proposoal of \cite{hms2}.}
\begin{equation}\begin{split}\label{solorig}
& 2\pi k^{-1} \widetilde{ds}{}^2 = \frac{1}{1+2b^2}\big[ d\tilde\xi^2 - \coth^2 \tilde \xi \, d\tilde t^2
+ 4 b^2(1+b^2)(\cosh \tilde t \, d\tilde \xi + \coth \tilde \xi \sinh \tilde t \, d\tilde t)^2
\\
& \hspace{100pt} + d \varphi^2 + \cot^2 \varphi \, d\zeta^2
+ 4 b^2(1+b^2)(\cos \zeta \, d\varphi + \cot \varphi \sin \zeta \, d\zeta)^2 \big] \ ,
\\
& e^{\widetilde{\Phi}} = \frac{e^{\widetilde{\Phi}_0}}{\sinh \tilde \xi\sin\varphi}\ ,
\\
& \sqrt{2\pi k^{-1}} \widetilde{A} = -
4 \sqrt{\frac{b^2 (b^2+1)}{1+ 2 b^2}} e^{-\widetilde{\Phi}_0} \big[ c_1 \cos\varphi\cos\zeta \, d(\cosh\tilde \xi\sinh \tilde t) + c_2 \cosh \tilde \xi \cosh \tilde t\, d(\cos\varphi \sin \zeta)\big]
\ .
\end{split}\end{equation}
Here the free constants $c_1$ and $c_2$ satisfy
\begin{equation}
c_1^2 + c_2^2 =1\ ,
\end{equation}
and encode the usual freedom of $U(1)$ electromagnetic duality rotations in 4d.
The choice $c_1=c_2= { 1 \ov \sqrt 2}$
ensures symmetry between the two coset factors.

Analytically continuing the $AdS_2$ coset part to the patch of interest \eqref{patch1}
\begin{equation}
\tilde \xi = i t \ , \qquad \tilde t = \xi \ , \qquad e^{\widetilde{\Phi}_0} = i e^{\Phi_0} \ ,
\end{equation}
gives the following solution of the equations of motion \eqref{seom}
supporting the metric \eqref{lag1}
\begin{equation}\begin{split}\label{solnew}
& 2\pi k^{-1}ds^2 = \frac{1}{1+2b^2}\big[ -dt^2 + \cot^2 t \, d\xi^2
- 4 b^2(1+b^2)(\cosh \xi \, dt - \cot t \sinh \xi \, d\xi)^2
\\
& \hspace{100pt} + d \varphi^2 + \cot^2 \varphi \, d\zeta^2
+ 4 b^2(1+b^2)(\cos \zeta \, d\varphi + \cot \varphi \sin \zeta \, d\zeta)^2 \big] \ ,
\\
& e^{\Phi} = \frac{e^{\Phi_0}}{\sin t\sin\varphi}\ ,
\\
& \sqrt{2\pi k^{-1}}A =
4 i \sqrt{\frac{b^2 (b^2+1)}{1+ 2 b^2}} e^{-\Phi_0} \big[ c_1 \cos\varphi\cos\zeta \, d(\cos t \sinh \xi) + c_2 \cos t \cosh \xi\, d(\cos\varphi \sin \zeta)\big]
\ .
\end{split}\end{equation}
The 1-form of the supergravity solution in \eqref{solorig} is real for real
$b$. The analytic continuation to this new patch leads to an imaginary 1-form
if $b$ is real. 

This raises an interesting question. If this background does
correspond to the $\l$-deformation \eqref{laghms} \cite{hms2} of the
superstring sigma model, then for some (perfectly legitimate) choices of the
$SO(1,2)$ gauge-fixed group field \eqref{patch1} we should end up with an
action that is not manifestly real. However, the reality of the action
\eqref{laghms} seems to follow in the usual way from considering the real form
of the superalgebra. The non-reality should only manifest itself in the
fermionic sector (as $i$ appears in the RR flux) and could arise from an
obstruction in the procedure of gauge-fixing the supergroup field of
\eqref{laghms} and integrating out the superalgebra-valued gauge field, but it
is not immediately clear why this should happen. At the same time, the
imaginary RR flux may be expected, given that \eqref{laghms} can be interpreted
as a deformation of the non-abelian T-dual of the $AdS_n \times S^n$ string
model with the duality applied to all space-time dimensions including time (cf.
\ci{hrt,Demulder:2015lva,Hull:1998vg}).
Note, however, that the gauge field in the action \rf{laghms} of the \lmo belongs to the 
superalgebra, and thus the non-abelian T-duality in \rf{9} is performed also in the fermionic directions
(cf. \ci{Berkovits:2008ic}), which may also have an effect on the issue of the reality 
of the corresponding RR flux. 

As here we are interested
in the special limit (and analytic continuation) \rf{part2} of the above background
combined with the analytic continuation of the parameters (i.e.
with $b$ and $k$ taken complex as in \rf{219},\rf{18})
we may formally consider the
solutions of the complexified theory, discussing the reality issue only
at the end.
It is worth recalling however, as discussed in section \ref{secpohlmeyer}, that if we 
analytically continue to $H^2 \times dS_2$ using \eqref{dsac},\eqref{hac},
while the background \eqref{lag1} still has an imaginary 1-form, the
special limits we consider below become real for real $b$
(as in \rf{bbb} compared to \rf{bb}).

The first limit we will take is as in \rf{part2} combined with infinite shift of the dilaton
\begin{align}\la{48}
& t\to t + \frac{i}{2} \log \big[\frac{1-\varkappa ^2 \rho ^2}{1+\rho ^2}\big]+i \log \gamma \ , &&
\xi \to \frac{1}{2} \log \big[-\frac{1-\varkappa \rho }{1+\varkappa \rho }\big]\ ,
&& \Phi_0 \to \Phi_0 + \log\big[-\frac{\gamma^2}4\big] \ ,\nonumber
\\ &\varphi \to \varphi + \frac{i}{2} \log \big[\frac{1+\varkappa ^2 r^2}{1-r^2}\big]+i \log \gamma \ , &&
\zeta \to \frac{i}{2} \log \big[-\frac{1+i \varkappa r}{1-i \varkappa r}\big] \ ,
&& \gamma \to \infty \ .
\end{align}
Starting from \rf{solnew} we then get the following
solution of the 4d supergravity equations \rf{seom} supporting the metric \rf{metdmv2} of the \emo
\begin{align}\nonumber
& 2h^{-1}ds^2 = -\frac{1+\rho^2}{1-\varkappa^2\rho^2} dt^2 + \frac{d\rho^2}{(1-\varkappa^2\rho^2)(1+\rho^2)}
+\frac{1-r^2}{1+\varkappa^2r^2}d\varphi^2 + \frac{dr^2}{(1+\varkappa^2r^2)(1-r^2)}\ ,
\\\nonumber
& e^\Phi =
e^{\Phi_0 + i (t+\varphi )} \frac{\sqrt{1+\rho ^2} \sqrt{1-r^2} }{\sqrt{1-\varkappa ^2 \rho ^2} \sqrt{1+\varkappa ^2 r^2}}\ ,
\\
& \sqrt{2h^{-1}}A =
\frac{2\sqrt{1+ \varkappa ^2} e^{-\Phi_0 -i (t+\varphi )}}{ \sqrt{1+\rho ^2} \sqrt{1-r^2}}
\big[c_1 r \, d\big(t-\frac i2 \log(1+\rho^2)\big) - c_2 \rho \, d\big(\varphi - \frac i2 \log(1-r^2)\big)\big] \ , \nonumber
\\ & \sqrt{2h^{-1}} e^\Phi F = - \frac{2\sqrt{1+\vk^2}}{\sqrt{1+\r^2}\sqrt{1-r^2}}\nonumber
\big[ c_1(e^0 \wedge e^3 - \r r\, e^1 \wedge e^2 
- i r \, e^0 \wedge e^2 -i \rho \, e^1 \wedge e^3) 
\\ & \hspace{150pt} + c_2( \r r \, e^0 \wedge e^3 + e^1 \wedge e^2 
- i \r \, e^0 \wedge e^2 + i r \, e^1 \wedge e^3) \big]\ , \label{sol1}
\end{align}
where we have defined the frame fields
\begin{equation*}
e^0 = \frac{\sqrt{1+\rho^2}}{\sqrt{1-\varkappa^2\rho^2}} dt \ , \quad
e^1 = \frac{d\rho}{\sqrt{1-\varkappa^2\rho^2}\sqrt{1+\rho^2}} \ , \quad
e^2 = \frac{\sqrt{1-r^2}}{\sqrt{1+\vk^2 r^2}} d\vp \ , \quad
e^3 = \frac{dr}{\sqrt{1+\varkappa^2r^2}\sqrt{1-r^2}} \ .
\end{equation*}
This background looks strange:
the $\varkappa \to 0$ limit of \eqref{sol1} gives the undeformed $AdS_2 \times S^2$ metric
supported by a non-trivial complex dilaton and RR flux
that explicitly depend on $t$ and $\varphi$.
While $t$ and $\varphi$ are still isometries of the metric and $ e^{\Phi}F$, which
enter the classical GS superstring action, the dilaton and RR 1-form are only invariant under
the combined transformation\foot{Formally the dilaton and RR 1-form are invariant under
separate shifts in $t$ and $\varphi$ if one is also allowed to shift $\Phi_0$.
Note also that the linear terms in the dilaton have
their origin in the large distance asymptotics of the background
corresponding to the gWZW model when the metric becomes flat
while the dilaton becomes linear, cf. \rf{solnew},\rf{48}.}
\begin{equation}\label{symtp}
t \to t + c \ , \qquad \varphi \to \varphi - c \ .
\end{equation}
This is different from the expected Bertotti-Robinson type flux supporting $AdS_2 \times S^2$.

If we instead consider the $\varkappa \to \infty$ limit of
\eqref{sol1}, as taken in \cite{mirror0}, i.e. first rescaling
\begin{equation}\label{kainflim}
t \to \varkappa^{-1} t \ , \qquad
\rho \to \varkappa^{-1} \rho \ , \qquad
\varphi \to \varkappa^{-1} \varphi \ , \qquad
r \to \varkappa^{-1} r \ , \qquad
h \to h \varkappa^2 \ ,
\end{equation}
we find the following real supergravity solution 
\begin{align}\label{solmirror}
&2h^{-1} ds^2 = -\frac{dt^2}{1-\rho^2} + \frac{d\rho^2}{1-\rho^2}
+\frac{d\varphi^2}{1+r^2} + \frac{dr^2}{1+r^2}\ , \hspace{50pt}
e^\Phi = \frac{e^{\Phi_0 } }{\sqrt{1- \rho ^2} \sqrt{1+ r^2}}\ .
\\\nonumber
&\sqrt{2h^{-1}} A = 2e^{-\Phi_0 } \big[c_1 r\, dt - c_2 \rho\, d\varphi \big]\ , \hspace{25pt}
\sqrt{2h^{-1}}e^{\Phi} F = - \frac{2}{\sqrt{1-\rho^2}\sqrt{1+r^2}} \big[
c_1 dt \wedge d r
-c_2 d\varphi \wedge d\rho \big]\ .
\end{align}
This is precisely the solution of the ``mirror'' model constructed in
\cite{mirror0} and is related to a $dS_2 \times H^2$ background by T-dualities
in $t$ and $\varphi$, giving an imaginary RR flux as might be expected (cf.
\cite{Hull:1998vg}).

\

The second limit we will consider is
\begin{align}\la{lim}
& t\to i \varkappa t + \frac{i}{2} \log \big[\frac{1+\rho^2}{1-\varkappa ^2 \rho ^2}\big]+i \log \gamma \ , &&
\xi \to \frac{1}{2} \log \big[\frac{1- i \rho }{1+i \rho }\big]\ ,
&& \Phi_0 \to \Phi_0 + \log\big[- \frac{\gamma^2}4\big] \ ,\nonumber
\\ &\varphi \to i\varkappa\varphi + \frac{i}{2} \log \big[\frac{1-r^2}{1+\varkappa ^2 r^2}\big]+i \log \gamma \ , &&
\zeta \to \frac{i}{2} \log \big[\frac{1- r}{1+r}\big] \ ,
&& \gamma \to \infty \ .
\end{align}
The resulting solution of \rf{seom} is given by
\begin{align}\nonumber
&2h^{-1} ds^2 = -\frac{1-\varkappa^2\rho^2}{1+\rho^2} dt^2 + \frac{d\rho^2}{(1-\varkappa^2\rho^2)(1+\rho^2)}
+\frac{1+\varkappa^2 r^2}{1-r^2}d\varphi^2 + \frac{dr^2}{(1+\varkappa^2r^2)(1-r^2)}\ ,
\\\nonumber
& e^\Phi =
e^{\Phi_0 - \varkappa (t+\varphi )} \frac{\sqrt{1-\varkappa^2\rho ^2} \sqrt{1+\varkappa^2 r^2} }{\sqrt{1+ \rho ^2} \sqrt{1- r^2}}\ ,
\\\nonumber
& \sqrt{2h^{-1}}A =
-\frac{2i\sqrt{1+ \varkappa ^2} e^{-\Phi_0 +\varkappa (t+\varphi )} }{\sqrt{1-\varkappa^2\rho ^2} \sqrt{1+\varkappa^2r^2}}
\big[c_1 \rho\, d\big(t-\frac1{2\varkappa} \log(\frac{1-\varkappa^2\rho^2}{\varkappa^2\rho^2})\big)
+ c_2 r\, d\big(\varphi -\frac{1}{2\varkappa} \log(\frac{1+\varkappa^2r^2}{\varkappa^2r^2})\big)\big] \ ,
\\ & \sqrt{2h^{-1}} e^\Phi F = - \frac{2i\sqrt{1+\vk^2}}{\sqrt{1-\vk^2\r^2}\sqrt{1+\vk^2r^2}}\nonumber
\big[ c_1(\vk^2 \r r\,e^0 \wedge e^3 - e^1 \wedge e^2 
- \vk \r \, e^0 \wedge e^2 + \vk r \, e^1 \wedge e^3) 
\\ & \hspace{170pt} + c_2( e^0 \wedge e^3 + \vk^2 \r r\, e^1 \wedge e^2 
+ \vk r \, e^0 \wedge e^2 + \vk \r \, e^1 \wedge e^3) \big]\ , \label{sol2}
\end{align}
where the frame fields are given by
\begin{equation*}
e^0 = \frac{\sqrt{1-\vk^2\rho^2}}{\sqrt{1+\rho^2}} dt \ , \quad
e^1 = \frac{d\rho}{\sqrt{1-\varkappa^2\rho^2}\sqrt{1+\rho^2}} \ , \quad
e^2 = \frac{\sqrt{1+\vk^2r^2}}{\sqrt{1- r^2}} d\vp \ , \quad
e^3 = \frac{dr}{\sqrt{1+\varkappa^2r^2}\sqrt{1-r^2}} \ .
\end{equation*}
There is a formal map between the two solutions \rf{sol1} and \rf{sol2} given by
\begin{equation}\label{mappp}
t \to i \varkappa t \ , \qquad \rho \to -\frac{i}{\varkappa\rho} \ , \qquad
\varphi \to i \varkappa \varphi \ , \qquad r \to \frac{i}{\varkappa r} \ .
\end{equation}
The metric of \eqref{sol2} is the double T-dual (in $t$ and $\varphi$) of the
metric of \eqref{sol1}. However, this T-duality relation does not obviously
extend to the full backgrounds as shifts in $t$ and $\varphi$ are not
isometries of the dilaton and the RR 1-form.\foot{It \label{foottd} may
still be possible to define a generalization of the T-duality rules that will
apply in the present situation. 
The dilaton coupling in the string action is given  by 
$\sqrt{-h}R^{(2)}\Phi = - 2\, \partial^2 \omega \, \Phi$ (in conformally flat
coordinates $h_{\a\b} = e^{2\omega} \eta_{\a\b}$).
Therefore, if $\Phi$ has a term linear in a target-space direction (which is
otherwise isometric, i.e. enters the string action only through its derivatives), we can integrate by parts and then perform the T-duality
transformation in the usual manner.  The resulting action will have a term
proportional to $(\partial \omega)^2$ whose role is to cancel the conformal
anomaly.
As the dilaton coupling  term is subleading in $\a'$
the T-dual classical superstring 
action can be found by the usual rules. One can then formally read off the corresponding
metric, $B$ field and $e^{\Phi}$ times the RR fluxes from the resulting action.
 They  need  not  by themselves satisfy the Type IIB supergravity equations of motion as these
follow from the vanishing of the one-loop Weyl anomaly beta-functions  and  thus are sensitive
to the full  dilaton coupling  and, in particular, the  central charge shift mentioned above. 
 The  resulting dilaton  of the T-dual   background may then be determined by solving these equations. }
Again they are only invariant under the combined transformation \eqref{symtp}.

The $\varkappa \to 0$
limit of \eqref{sol2} is much simpler
than that of \rf{sol1}\foot{The apparent divergence of the RR potential
turns out to be a total derivative and can therefore be removed by an appropriate gauge
transformation
\begin{equation*}\begin{split}
\sqrt{2h^{-1}}A& \to \sqrt{2h^{-1}}A + d\big(\frac{2i\sqrt{1+ \varkappa ^2} e^{-\Phi_0 +\varkappa (t+\varphi )} }{\varkappa\sqrt{1-\varkappa^2\rho ^2} \sqrt{1+\varkappa^2r^2}}
(c_1 \rho +c_2 r )\big)
\\ & =
\frac{2i\sqrt{1+ \varkappa ^2} e^{-\Phi_0 +\varkappa (t+\varphi )}}{\sqrt{1-\varkappa^2\rho ^2} \sqrt{1+\varkappa^2r^2}}
\big[ c_1 \rho \, d\big(\varphi - \frac{1}{2\varkappa}\log(1+\varkappa^2r^2)\big)
+c_2 r \, d\big(t-\frac{1}{2\varkappa}\log(1-\varkappa^2\rho^2)\big)
\big]\ .
\end{split}\end{equation*}}
\begin{align}\label{sol3}
&2h^{-1} ds^2 = -\frac{dt^2}{1+\rho^2} + \frac{d\rho^2}{1+\rho^2}
+\frac{d\varphi^2}{1-r^2} + \frac{dr^2}{1-r^2}\ , \hspace{50pt}
e^\Phi = \frac{e^{\Phi_0 } }{\sqrt{1+ \rho ^2} \sqrt{1- r^2}}\ .
\\\nonumber
&\sqrt{2h^{-1}} A = 2ie^{-\Phi_0 } \big[c_1 \rho \, d\varphi + c_2 r\, dt \big]\ , \hspace{25pt}
\sqrt{2h^{-1}}e^{\Phi} F = - \frac{2i}{\sqrt{1+\rho^2}\sqrt{1-r^2}} \big[
c_1 d\varphi \wedge d \rho
+c_2 dt \wedge dr \big]\ .
\end{align}
Performing T-dualities in both $t$ and $\varphi$ we recover
the standard Bertotti-Robinson solution with constant dilaton and homogeneous RR flux:
\begin{align}\label{sol4}
& 2h^{-1}ds^2 = -(1+\rho^2)dt^2 + \frac{d\rho^2}{1+\rho^2}
+ (1-r^2)d\varphi^2 + \frac{dr^2}{1-r^2}\ , \hspace{50pt}
e^\Phi = e^{\Phi_0 } \ ,
\\\nonumber
& \sqrt{2h^{-1}} A = 2e^{-\Phi_0 } \big[c_1 \rho\, dt - c_2 r\, d\varphi \big]\ , \hspace{25pt}
\sqrt{2h^{-1}}e^{\Phi} F = - 2\big[ c_1 dt \wedge d\rho - c_2 d\varphi \wedge dr\big]\ .
\end{align}
This suggests that if the metric and $e^\Phi F$ of the solution \rf{sol2} can
be formally T-dualized for $\varkappa \neq 0$ (e.g. by applying the standard
T-duality rules to just these combinations of the background fields, see
footnote \ref{foottd}) it will give a real ``background'' for the metric
\rf{metdmv2} (the T-duality in $t$ will remove the factor of $i$ in $F$).  It
would be interesting to see if this bears any relation to the
$\eta$-deformation \eqref{lagdmv} of the $AdS_2 \times S^2$ supercoset model.
Having a factorized (but not isometric) dilaton, this background will be
obviously different from the solution constructed in 
\ci{Lunin:2014tsa}\foot{In \ci{Lunin:2014tsa} the independence of the 
dilaton and RR fields from the isometric directions of the metric was
assumed from the start.} and its
meaning remains to be understood.
Finally, given that the standard
Bertotti-Robinson solution appears (after T-dualities) in the $\varkappa \to 0$
limit of \eqref{sol2}, while the ``mirror'' model \eqref{solmirror} of
\cite{mirror0} appears in the $\varkappa \to \infty$ limit of \eqref{sol1}, it
would be interesting to see if the map \eqref{mappp} between the two
backgrounds \eqref{sol1},\eqref{sol2} is related to the ``mirror duality'' of
\cite{mirror0,mirror}.

\

Finally, let us note that the $\varkappa \to i$ limit of \rf{sol1} or \rf{sol2} can be taken
as in \rf{111}\foot{One can also use \eqref{111a}
\begin{equation*}
t = \epsilon \xxp + \frac{\xxm}{\epsilon} \ , \qquad
\varphi = \epsilon \xxp - \frac{\xxm}{\epsilon} \ , \qquad
\rho = \cot \alpha \ , \qquad r = \coth\beta \ , \qquad \varkappa = \sqrt{-1 - s\, \epsilon^2}\ .
\end{equation*}
leading to the same pp-wave type background. This is a consequence of the formal map \eqref{mappp} between
\eqref{sol1} and \eqref{sol2}.}
\begin{equation}\la{420}
t = \epsilon \xxp + \frac{\xxm}{\epsilon} \ , \qquad
\varphi = \epsilon \xxp - \frac{\xxm}{\epsilon} \ , \qquad
\rho = \tan \alpha \ , \qquad r = \tanh\beta \ , \qquad \varkappa = \sqrt{-1 + s\, \epsilon^2}\ .
\end{equation}
Choosing $s = 1$ for the solution \eqref{sol1} and $s = -1$ for
\eqref{sol2} and then sending $\epsilon \to 0$, in
both cases we find the following pp-wave background
\begin{equation}\begin{split}
& 2h^{-1}ds^2 = -4d\xxp d\xxm + \frac12(\cos 2\alpha - \cosh 2\beta)(d\xxm)^2 + d\alpha^2 + d\beta^2 \ ,
\\
& e^\Phi = e^{\Phi_0}\ , \qquad
\sqrt{2h^{-1}}A = 2e^{-\Phi_0}\big[ \tilde c_1\cos\alpha\sinh\beta + \tilde c_2\sin\alpha\cosh\beta\big] d\xxm\ ,
\end{split}\end{equation}
where $\tilde c_{1,2} = \pm c_{1,2}$.
This is the pp-wave background of \cite{hrt}, whose light-cone gauge-fixing ($\xxm = \mu \tau$)
yields the Pohlmeyer-reduced theory for $AdS_2 \times S^2$,
equivalent \ci{Grigoriev:2007bu} to the $\mathcal{N}=2$
supersymmetric sine-Gordon model.
If we had taken the opposite signs for $s$ in \rf{420} we would have ended
up with the same solution with $\xxm \to i \xxm$. The light-cone gauge-fixing
of this model gives the Pohlmeyer-reduced theory for $H^2 \times dS^2$.

Let us also note that if we set $\varkappa = i$ in the solutions \eqref{sol1}
and \eqref{sol2} without the rescaling of $x^\pm$ in \eqref{420} we find a
simple string background given by a flat metric with vanishing RR 1-form and a
dilaton linear in the null direction $t + \varphi$ (the factor of $\pm i$ in
the dilaton can be removed by a simple analytic continuation of $t$ and
$\varphi$).

\subsection{Algebraic coordinates}

The \lmo solutions \eqref{solorig} and \eqref{solnew} take 
remarkably simple forms in the algebraic coordinates introduced in \rf{311},\rf{312}.
The solution \eqref{solnew} becomes\foot{This form of the solution
manifestly realizes the observation of \cite{Demulder:2015lva} that the $\lambda$-deformation
amounts to rescaling the tangent space directions of the gauged WZW model for $F/G$
(here $\frac{SO(1,2)}{SO(1,1)} \times \frac{SO(3)}{SO(2)}$, given by the point $b=0$)
while leaving the dilaton invariant 
and with the RR flux depending on the deformation parameter only through an overall constant factor.}
\begin{equation}\begin{split}\label{solpq}
&2\pi k^{-1} ds^2 = \frac1{1-x^2+y^2}\big[-(1+2b^2) dx^2 + \frac{dy^2}{1+2b^2} \big]
+ \frac1{1-p^2-q^2}\big[(1+2b^2) dp^2 + \frac{dq^2}{1+2b^2}\big] \ ,
\\ & e^\Phi = \frac{e^{\Phi_0}}{\sqrt{1-x^2+y^2}\sqrt{1-p^2-q^2}} \ , \qquad
\sqrt{2\pi k^{-1}}A = 4i \sqrt{\frac{b^2(b^2+1)}{1+2b^2}} e^{-\Phi_0}\big[c_1 p\, dy + c_2 x\, dq \big] \ .
\end{split}\end{equation}
Note that a formal analytic continuation of this background by setting $x= i y', \ y= i x'$ gives a real solution 
\begin{equation}\begin{split}\label{solpqr}
&2\pi k^{-1} ds^2 = \frac1{1-x'^2+y'^2}\big[-\frac{dx'^2}{1+2b^2} + (1+ 2 b^2) {dy'^2} \big]
+ \frac1{1-p^2-q^2}\big[(1+2b^2) dp^2 + \frac{dq^2}{1+2b^2}\big] \ ,
\\ & e^\Phi = \frac{e^{\Phi_0}}{\sqrt{1-x'^2+y'^2}\sqrt{1-p^2-q^2}} \ , \qquad
\sqrt{2\pi k^{-1}}A = - 4 \sqrt{\frac{b^2(b^2+1)}{1+2b^2}} e^{-\Phi_0}\big[c_1 p\, dx' + c_2 y'\, dq \big] \ .
\end{split}\end{equation}
If instead we formally continue \rf{solpq} to the region for which $x^2 - y^2 >
1$, we find (after setting $ e^{\widetilde{\Phi}_0} = i e^{\Phi_0}$) a
different real background, which represents the solution \eqref{solorig}, i.e.
the original solution of \cite{Sfetsos:2014cea} corresponding to the metric in
the coordinate patch in \rf{lag1a}.

Using the relations \rf{219},\rf{18} between the parameters
and taking the scaling limit \rf{315} combined with a redefinition of the
dilaton $e^{\Phi_0} \to i \g^2 e^{\Phi_0}$
the solution \eqref{solpq} becomes simply
\begin{equation}\begin{split}\label{solpqlimit}
& 2h^{-1} ds^2 = \frac1{y^2- \varkappa^2x^2}\big( dx^2 + dy^2 \big)
+ \frac1{q^2 + \varkappa^2p^2}\big( -dq^2 + dp^2 \big) \ ,
\\ & e^\Phi = \frac{e^{\Phi_0}}{\sqrt{y^2 - \varkappa^2x^2}\sqrt{q^2 + \varkappa^2 p^2}} \ , \qquad
\sqrt{2h^{-1}} A = -2i \sqrt{1+\varkappa^2} \, e^{-\Phi_0}\, \big(c_1 p\, dy +c_2 x\, dq \big)\ .
\end{split}\end{equation}
One can check directly that the supergravity equations of motion \rf{seom} are
indeed satisfied.\foot{To recall, $c_1$ and $c_2$  are arbitrary constants
satisfying $c_1^2  + c_2^2 =1$, so a  symmetric choice is  $c_1=c_2 = {1 \ov
\sqrt 2}$.} This solution may be interpreted as a deformation of a $H^2 \times
dS_2$ background (for which an imaginary RR flux could be expected, cf.
\ci{Hull:1998vg}).  For $\varkappa=0$ the dilaton is non-constant but it can be
eliminated by T-dualities in the $x$ and $p$ directions, which along with
sending $y\to y^{-1}$ and $q\to q^{-1}$ leaves the metric invariant.

The metric and $e^\Phi F$ of \eqref{solpqlimit} are invariant under
separate rescalings of $(x,y)$ and $(p,q)$, however, as discussed above
the dilaton and RR 1-form are only invariant when these rescalings are correlated
as $(x,y) \to e^{\tilde{c}} (x,y)$, $(p,q) \to e^{-\tilde{c}} (p,q)$,
which corresponds to the symmetry \eqref{symtp} of the backgrounds \eqref{sol1},\eqref{sol2}.

A similar background representing the  deformation of $AdS_2 \times S^2$ may
be found using a different real slice of the diagonal coordinates as
in \rf{a8}. Setting 
\be 
&&  y= e^{i \vp}    \cosh v \ , \ \ \ \ x= i e^{i \vp}    \sinh v \ , \ \ \ \    r= \tanh v   \  , \no  \\
&& q= e^{i t}    \cos \a \ , \ \ \ \  \ \  \ p= i e^{i t}    \sin \a \ , \ \ \ \  \  \  \rho = \tan \a   \  ,  \la{cy} 
\ee
we find that \rf{solpqlimit} then transforms into the background \rf{sol1} found earlier. 

\section{Poisson-Lie duality interpretation}\label{sec7}

Apart from the relation between the \lmo and \emo through a scaling limit and analytic continuation
described in section \ref{sec2}, which is somewhat unexpected (though partly prompted by the
natural map between the parameters \rf{219},\rf{18},\rf{19}),
one may anticipate that the two models may be in some sense dual to each other.
Indeed, the undeformed limit of the \emo is the standard supercoset model, while
the undeformed limit of the \lmo is the non-abelian T-dual of the latter (cf. \rf{9},\rf{17}).
A natural suggestion is then that the two models may be related by the Poisson-Lie (PL) duality
of \ci{Klimcik:1995ux,Klimcik:1996np}.

Below we will directly verify this conjecture on the simplest example of the bosonic $S^2$ coset.
The corresponding metric of the \lmo is in the second line of \rf{lag1} (or, in diagonal form, the second term of \rf{312}),
and its \emo counterpart is in the second line of \rf{metdmv2}.
We are going to compare them with the PL dual pair of models associated to the $SL(2,\mathbb{C})$
double \ci{Klimcik:1996np,Sfetsos:1999zm}:
the first corresponds to the $SU(2)$ subgroup and the second to the Borel subgroup $B_2$
(upper triangular matrices with reals on diagonal). 
The corresponding metrics are given, e.g., in equations 3.18 and 3.19
of \cite{Sfetsos:1999zm} with two free parameters $ \aa,\bb$
and with an overall coefficient $\text{\hhh}$.\foot{We denote the
parameters $a,b$ of \cite{Klimcik:1996np,Sfetsos:1999zm} by roman letters.}

The first metric is
\begin{equation} \la{1st}
ds^2_1=\frac{ \hhh
\, \aa}{\text{a}^2 + (\text{b}-\cos \theta)^2 } (d\theta^2+ \sin^2 \theta d\varphi^2) \ .
\end{equation}
Setting $\text{b}=0$ (which is required to get the integrable model we are interested in here)
and
\begin{equation}\label{ta1}
\text{\hhh} = \frac{h}{2\varkappa} \ ,\qquad\qquad \text{a}=\varkappa^{-1}\ ,
\end{equation}
we find that \rf{1st} becomes precisely
the corresponding \emo metric in \rf{metdmv2} (where $r=\cos\theta$).

The second metric of the PL dual pair is \cite{Sfetsos:1999zm}
\begin{equation}
ds^2_2= \frac{ \text{\hhh}\, \text{a}_1}{2(1+\text{a}_1 z)} \Big(\frac{dz^2}{\rho^2} + \big[d\rho+\big(\frac{\text{b}-1}{\text{a}}+\frac{z-\frac{\text{a}_1} {4}\rho ^2}{1+ \text{a}_1 z}\big)\frac{dz}{\rho }\big]^2\Big) \ , \qquad \text{a}_1 \equiv \frac{2\text{a}}{\text{a}^2 +(\text{b}-1)^2} \ .
\end{equation}
Setting $\text{b}=0$ and doing a field redefinition
to put this metric into a diagonal form
\begin{equation}
z = \frac12(\text{a}+\text{a}^{-1})\big[ (p+q)^2-1\big] \ , \qquad
\qquad \rho = (\text{a}+\text{a}^{-1})\sqrt{p^2-q^2 -1}\ ,
\end{equation}
we find
\begin{equation}\label{mm} ds^2_2=
\frac{\text{\hhh}}{p^2-q^2-1}\big(\, \text{a}\, dp^2 + \text{a}^{-1}dq^2 \big)\ .
\end{equation}
Making further redefinitions
\begin{equation}\label{ta2}
\text{\hhh} = \frac{k}{2i\pi} \ ,\qquad \qquad \text{a} = -i(1+2b^2) \ ,\qquad \qquad q\to i q \ ,
\end{equation}
we obtain the metric of the $\l$-deformation of the non-abelian T-dual of $S^2$ in the algebraic
coordinates used in \rf{311},\rf{312},\eqref{solpq}
\begin{equation} ds^2=
\frac{k}{2\pi}\frac{1}{1-p^2-q^2}\big[(1+2b^2)dp^2 + \frac{dq^2}{1+2b^2}\big]\ .
\end{equation}
Note that the definitions in \eqref{ta1} and \eqref{ta2} are related by the map
\eqref{219},\rf{18} precisely as required by our general discussion in sections \ref{sec1} and \ref{sec21}.

This implies that in the $S^2$ coset case, the $\eta$-deformation of
\cite{Delduc:2013fga} is Poisson-Lie dual to an analytic continuation of the
$\l$-deformation of \cite{Sfetsos:2013wia,hms1}. A similar relation should
then be expected in general.

\section*{Acknowledgments}

We would like to thank J.L. Miramontes for important discussions and for
sharing with us an unpublished draft on the derivation of the Pohlmeyer reduced
model from from $\l$-deformed action that partially motivated our construction
of the limit discussed in section \ref{sec2}. We also thank O. Lunin and R.
Roiban for useful comments on the draft and F. Delduc, M. Magro, K. Sfetsos, D.
Thompson, S. van Tongeren, B. Vicedo and L. Wulff for discussions.
The work of BH was funded by the DFG through the Emmy Noether Program ``Gauge
Fields from Strings'' and SFB 647 ``Space - Time - Matter. Analytic and
Geometric Structures.''
The work of AAT was supported by the ERC Advanced grant No.290456,
STFC Consolidated grant ST/J0003533/1 and RNF grant
14-42-00047.

While this paper was in preparation there appeared ref. \ci{Vicedo:2015pna} giving
a general Hamiltonian construction of the relation of the two deformed models via the
Poisson-Lie duality complementing our discussion in section \ref{sec7}.

\appendix

\section{Different forms of deformed metrics in \texpdf{$SO(3)/SO(2)$}{SO(3)/SO(2)} case} \label{appb}
\def\theequation{A.\arabic{equation}}
\setcounter{equation}{0}

The $\lambda$-deformed metric \ci{Sfetsos:2013wia} corresponding to the $S^2$ coset (given in the second line of \rf{lag1})
can be written, after a simple
change of coordinates, $z= \cos \zeta, \ w= \cos \vp\, \sin \zeta$, in the following form
(ignoring overall factors)
\be \la{a1}
ds^2= {1\ov 1-z^2} \big[ dz^2 +{ z^2\ov 1-w^2} dw^2 + m^2 ( w dz + z dw )^2 \big]
\ , \qquad \qquad m^2\equiv 4b^2(1+b^2) = - \k^{-2} -1 \ .\ee
The non-abelian T-dual of $S^2$ is found in the limit $m \to \infty$ with
$z=1 - {1 \ov 2m^2} Z^2, \ w= 1 - {1 \ov 2m^2} W^2$ giving \ci{Klimcik:1996np,bal} \
$ds^2= Z^{-2} ( d W^2 + {1 \ov 4} [d (W^2 + Z^2)]^2 )$.

Introducing the new coordinates $X, Y$ and $P,Q$ as
\be && e^Y = z \sqrt{1 + m^2 w^2} \ , \qquad \ \ \ \ \ \cos X = \sqrt{1-w^2\ov 1+ m^2 z^2} \la{a2} \ , \\
&& P= e^Y \cos X = z \sqrt{1 - w^2}\ , \ \ \ \ Q= e^Y \sin X = \sqrt{1+m^2} \, z w \ , \ \ \ \ \
P+i Q = e^{Y + i X} \ , \la{a3}
\ee
we can put \rf{a1} into the conformally-flat form (cf. \rf{312})
\be && ds^2 = { 1\ov e^{-2 Y} - { 1 \ov 1 + m^2} (1 + m^2 \cos^2 X)} ( dX^2 + dY^2) \la{a4} \\
&&\ \ \ \ \ = {1 \ov 1 - P^2 - {1 \ov 1 + m^2} Q^2 } (dP^2 + dQ^2) \ . \la{a5}
\ee
Here the $m=0$ limit corresponds to the $SO(3)/SO(2)$ gauged WZW metric.\foot{The
curvature of \rf{a5} is (setting $1+m^2 = -\k^{-2}$):
$\ R =-2 { 1 - \k^2 + (1+ \k^2) ( P^2 + \k^2 Q^2) \over 1 - P^2 + \k^2 Q^2 }$.}

One option to take a limit of this metric is to do an infinite rescaling of $P$ and $Q$ (combined with
the replacement of $m$ by $\k$ as in \rf{a1}), i.e.
to drop the constant $1$ in \rf{a5} (and reverse overall sign of the metric).
This leads to a scale-invariant (i.e. it has an isometry) metric as in \rf{315},\rf{316}
that is a deformation of $H^2$
\be \la{a6}
ds^2= {1 \ov P^2 - \k^2 Q^2 } (dP^2 + dQ^2) \ . \ee
Alternatively, we may consider the first form of the metric \rf{a4}
and set $P+ i Q = \exp( Y + i X) = \g \exp( i U - V)$
\be\la{a7}
Y=\ln \g + i U\ , \ \ \ \ X= i V \ , \ \ \ {\rm i.e.} \ \ \ \ \ P=\g\, e^{i U} \cosh V \ , \ \ \ \
Q = i \g\, e^{i U} \sinh V\ , \ \ \ \ \ \ \ \ \ \g \to \infty \ ,
\ee
i.e. use a different real slice where $U,V$ are real while $P,Q$ are not.
Then the $e^{-2 Y}$ term in \rf{a4} drops out and we find
\be \la{a8}
ds^2 = { 1 \ov \cosh^2 V + \k^2 \sinh^2 V} (dU^2 + dV^2) \ .
\ee
This is, indeed, the metric of the $\eta$-deformed $S^2$ space,\foot{The
sphere metric may be written as
${ 1 \ov \cosh^2 V} (dV^2 + dU^2) = d\a^2 + \cos^2 \a\, d U^2,$ \
$\tan {\a\ov 2} = \tanh { V\ov 2} $.}
i.e. it is equivalent to the second line of \rf{metdmv2} 
($\vp=U, \ r= \tanh V$)
\ci{Fateev:1992tk,Delduc:2013fga,hrt}.

A similar discussion can be repeated for the $AdS_2$ coset part of \rf{lag1}, obtaining
the first line of \rf{metdmv2} in the limit. 

\section{An alternative dilaton for the deformed models: \texpdf{$AdS_2 \times S^2$}{AdS2 x S2}} \label{appdil}
\def\theequation{B.\arabic{equation}}
\setcounter{equation}{0}

The dilaton discussed in section \ref{secsugra} (see \rf{solorig},\rf{solnew})
is the one  assumed as a starting point  for constructing supergravity
solutions for the \lmo  in  \cite{Sfetsos:2014cea,Demulder:2015lva}  and 
originates from integrating out 
the  gauge field $A_\pm $ (see \cite{tse} and references there) in the bosonic truncation \rf{16} of \rf{laghms}, i.e. 
\begin{equation}\label{det}
e^{2\Phi} = 
\frac1{\det\big[(\operatorname{Ad}_f - 1 - (\lambda^{-2} - 1) P_\lambda)|_{\hat{\mathfrak{f}}_0 \oplus \hat{\mathfrak{f}}_2}\big]} \ ,
\end{equation}
where the operator under the determinant is restricted to act on the
bosonic subalgebra of the superalgebra $\hat{\mathfrak{f}}$ and $f$ is
taken to be a bosonic coset representative.

For the \lmo associated to $AdS_2 \times S^2$ this gives the dilaton 
in \eqref{solorig} for the coset
representative \eqref{patch2}. For the coset representative
\eqref{patch1} we find the dilaton in \eqref{solnew}, i.e. 
\begin{equation}\label{bosdil}
e^{\Phi} = \frac{e^{\Phi_0}}{\sin t \sin \vp}\ .
\end{equation}

In \cite{hms2} an alternative expression for the dilaton was  proposed, which is
given by the superdeterminant arising from integrating out the complete
gauge field in \eqref{laghms}
\begin{equation}\label{sdet}
e^{2\Phi} = 
\frac{1}{\operatorname{sdet}\big[(\operatorname{Ad}_f - 1 - (\lambda^{-2} - 1) P_\lambda)\big|_{\hat{\mathfrak{f}}}\big]} \ , 
\end{equation}
where  now  the operator under the superdeterminant acts on the full superalgebra
$\hat{\mathfrak{f}}$.
As we are interested in the  bosonic supergravity background, the 
group field $f$  may  still   be  taken to be a bosonic coset
representative. Then  the operator under the superdeterminant
factorizes and \eqref{sdet} can be written as
\begin{equation}\label{sdeta}
e^{2\Phi} =
\frac{\det\big[(\operatorname{Ad}_f - 1 - (\lambda^{-2} - 1) P_\lambda)\big|_{\hat{\mathfrak{f}}_1 \oplus \hat{\mathfrak{f}}_3}\big]}
{\det\big[(\operatorname{Ad}_f - 1 - (\lambda^{-2} - 1) P_\lambda)\big|_{\hat{\mathfrak{f}}_0 \oplus \hat{\mathfrak{f}}_2}\big]}\ .
\end{equation}
The  denominator   factor  of \eqref{sdeta} is identical to \eqref{det} and therefore for the
\lmo associated to $AdS_2 \times S^2$ its contribution to the dilaton (for the
coset representative \eqref{patch1}) is again given by \eqref{bosdil}. 

 To compute the  contribution of the  fermionic 
numerator  factor   we need to consider the full superalgebra in
\eqref{1},\eqref{laghms} and not just its bosonic truncation. 
Starting with the
superalgebra $\mathfrak{psu}(1,1|2)$,\foot{We use the matrix representation
of $\mathfrak{psu}(1,1|2)$ given in appendix C of \cite{hrt}.}
which has the bosonic
subalgebra $\mathfrak {so}(1,2) \oplus \mathfrak{so}(3)$ required for
the $AdS_2 \times S^2$ case (the bosonic gauge group in \eqref{1} remains unchanged), we find the
contribution of the numerator of \eqref{sdeta} to $e^\Phi$  to be 
\begin{equation}
\begin{split}
& (1+\lambda^4 + 2  \lambda^2 \cosh 2 \xi )\cos^2 t 
+ (1+ \lambda^4 + 2\lambda^2 \cos 2 \zeta ) \cos^2\varphi   - (1-\lambda ^2)^2
\\ & \hspace{165pt}-4 \lambda  (1+\lambda ^2) \cos t \cos\vp  \cosh \xi  \cos \zeta\ .
\end{split}
\end{equation}
Combining this expression with \eqref{bosdil} we  arrive at  the following
(alternative  to \rf{bosdil})   proposal for the dilaton
\begin{equation}\begin{split}\label{sdil}
e^{\Phi} =  \frac{e^{\Phi_0}}{\sin t \sin \vp}  &
 \Big[(1+\lambda^4 + 2  \lambda^2 \cosh 2 \xi )\cos^2 t 
+ (1+ \lambda^4 + 2\lambda^2 \cos 2 \zeta ) \cos^2\varphi   - (1-\lambda ^2)^2
\\ &\hspace{169pt} -4 \lambda  (1+\lambda ^2) \cos t \cos\vp  \cosh \xi  \cos \zeta
\Big]\ .
\end{split}
\end{equation}
One can indeed check that together with the metric of \eqref{solnew} this
solves the dilaton equation, i.e. the first equation of \eqref{seom} as well as
the trace of the Einstein equation (the second equation of
\eqref{seom}).

The remaining equations involving RR   flux are no longer satisfied, i.e.  the
RR background needs to be modified.  How this should be done is not clear, but
it is worth noting that as the trace of the Einstein equation in \eqref{seom}
is still satisfied, the simplest consistent ansatz is for only  a single RR
1-form potential to be non-zero.\foot{One can try some simple ansatzes, such as
using the same RR 1-form as in \eqref{solnew}, or,  alternatively,  demanding
that $e^\Phi F$ is unchanged,  but  neither of these proposals work.} 

Let us note that in the algebraic coordinates
\eqref{311},\eqref{312} the dilaton \eqref{sdil} is given by
\begin{equation}\label{sdilpq}
e^{\Phi} = e^{\Phi_0}\frac{(1+\lambda ^2)^2 (x^2+p^2) - 4 \lambda (1+\lambda ^2) xp-(1-\lambda ^2)^2 (1+ y^2 -q^2)}{\sqrt{1-p^2-q^2} \sqrt{1-x^2+y^2}} \ .
\end{equation}
Here the denominator is the contribution from the bosonic sector \eqref{det},
i.e. the dilaton considered earlier in \eqref{solpq}.  Again one can check that
together with the metric of \eqref{solpq} this expression \eqref{sdilpq} solves
the dilaton equation and the trace of the Einstein equation.

Now let us take the two special limits \eqref{48} and \eqref{lim} of the new
dilaton \eqref{sdil}  (note that  here we will no longer need the infinite
shift of the constant part of the dilaton).  This leads to
\begin{equation}\label{sdil1}
e^\Phi = e^{\Phi_0} 
\frac{\sqrt{1+\rho ^2} \sqrt{1-r^2} \cos (t-\varphi )+i \sqrt{1+\varkappa ^2} \rho  r}{\sqrt{1-\varkappa ^2 \rho ^2} \sqrt{1+\varkappa ^2 r^2}}
\end{equation}
for the limit \eqref{48}, relating to the metric in \eqref{sol1}, and to 
\begin{equation}\label{sdil2}
e^\Phi = e^{\Phi_0}
\frac{\sqrt{1-\varkappa ^2 \rho ^2} \sqrt{1+\varkappa ^2 r^2} \cosh[ \varkappa  (t-\varphi )]+i \varkappa  \sqrt{1+\varkappa ^2} \rho  r}{\sqrt{1+\rho ^2} \sqrt{1-r^2}}
\end{equation}
for the limit \eqref{lim}, relating to the metric in \eqref{sol2}.  In the
$\varkappa \to \infty$ limit of \eqref{sdil1} (using \eqref{kainflim}) we
recover the dilaton of the ``mirror'' model \eqref{solmirror}, while taking the
$\varkappa \to 0$ limit of \eqref{sdil2} we recover the T-dual of the dilaton
of the background \eqref{sol3}.  Furthermore, we can recover the dilatons of
\eqref{sol1} and \eqref{sol2} from \eqref{sdil1} and \eqref{sdil2} respectively
(up to trivial signs) via an additional  infinite  constant shift of $t - \vp$
(along with compensating shifts of the constant part of the dilaton).
Equivalently, the expressions in  \eqref{sol1} and \eqref{sol2} can be  found
directly from \eqref{sdil} by decorrelating the limits  in the $AdS_2$ and
$S^2$ $\lambda$-models, i.e. using two separate parameters $\g$  for $t$ and
$\varphi$  in \eqref{48}  or \eqref{lim}. 

For  $\vk = i$,  when the metrics of \eqref{sol1} and \eqref{sol2} become flat,
any  ``null''  dilaton $e^{\Phi} = F(t\pm\vp)$  solves the dilaton equation and
the trace of the Einstein equation in \eqref{seom}.  Indeed, for $\vk=i$ the
dilatons \eqref{sdil1},\eqref{sdil2} take  this form.
Further, if we take $\vk = i$ without rescaling the coordinates, so that the
metric is Ricci flat, then asking that the RR fluxes vanish implies that
$e^{\Phi}$ is also a linear function of $t\pm \vp$.

Let us note that  the  dilatons \eqref{sdil1},\eqref{sdil2} are complex, so
their interpretation as part of  supergravity   solutions is unclear.  Also,
with the dilatons \eqref{sdil1},\eqref{sdil2}  having non-trivial (non-linear)
dependence   on  $t$ and $\vp$  the resulting  background    would be truly
non-isometric (with no  chance of  simplifying  T-duality transform).  This
suggests  that to recover the \emo from the \lmo we should  indeed consider the
decorrelated limit of \eqref{sdil} (with two separate infinite  $\g$
parameters), leading again to  the solutions \eqref{sol1} and \eqref{sol2}, for
which the dilatons {\em are} linear in $t\pm\vp$.



\end{document}